\documentclass[preprint,onecolumn,aps,prd,groupedaddress,nofootinbib,showpacs]{revtex4}
\usepackage{graphicx}
\DeclareGraphicsExtensions{.eps}
\usepackage{amsmath}
\usepackage{bm}
\usepackage{slashed}
\usepackage{epsfig}
\usepackage[hypertex]{hyperref}
\newcommand{\ben}{\begin{eqnarray}}
\newcommand{\een}{\end{eqnarray}}

\newcommand{\bef}{\begin{figure}[!htp]}
\newcommand{\eef}{\end{figure}}

\newcommand{\bea}{\begin{eqnarray}}
\newcommand{\eea}{\end{eqnarray}}

\newcommand{\state}[4]{{^#1\hspace{-0.6mm}#2_{#3}^{[#4]}}}

\newcommand\CScSa{\state{3}{S}{1}{1}}

\newcommand\COaSz{\state{1}{S}{0}{8}}

\newcommand\COcSa{\state{3}{S}{1}{8}}

\newcommand\COcPj{\state{3}{P}{J}{8}}

\begin{document}

\preprint{YITP-SB-14-43}

\title{Heavy Quarkonium Production at Collider Energies:\\ 
Partonic Cross Section and Polarization}

\author{Zhong-Bo Kang$^1$, Yan-Qing Ma$^{2,3}$,
Jian-Wei Qiu$^{4,5}$, and 
George Sterman$^{5}$}

\email{zkang@lanl.gov, yqma@bnl.gov, jqiu@bnl.gov, sterman@insti.physics.sunysb.edu}

\affiliation{$^1$Theoretical Division, 
                  Los Alamos National Laboratory, 
                  Los Alamos, NM 87545, USA}
                
\affiliation{$^2$Maryland Center for Fundamental Physics, 
		University of Maryland, College Park, MD 20742, USA}

\affiliation{$^3$Center for High-Energy Physics, 
		Peking University, Beijing, 100871, China}

\affiliation{$^4$Physics Department,
                Brookhaven National Laboratory,
                Upton, NY 11973-5000, USA}

\affiliation{$^5$C.N.\ Yang Institute for Theoretical Physics
             and Department of Physics and Astronomy,
             Stony Brook University, 
             Stony Brook, NY 11794-3840, USA}

\date{\today}

\begin{abstract}
We calculate the ${\cal O}(\alpha_s^3)$ short-distance, QCD collinear-factorized coefficient functions for all partonic channels that include the production of a heavy quark pair at short distances.   This provides the first power correction to the collinear-factorized inclusive hadronic production of heavy quarkonia at large transverse momentum, $p_T$, including the full leading-order perturbative contributions to the production of heavy quark pairs in all color and spin states employed in NRQCD treatments of this process.  We discuss the role of the first power correction in the production rates and the polarizations of heavy quarkonia in high energy hadronic collisions.  The consistency of QCD collinear factorization and non-relativistic QCD factorization applied to heavy quarkonium production is also discussed.
\end{abstract}

\maketitle

\section{Introduction}
\label{sec:intro}

Both the cross section and polarization of heavy quarkonium production in high energy collisions 
have posed significant challenges to our understanding of the production mechanism \cite{Brambilla:2010cs}. 
Although Non-Relativistic QCD (NRQCD) treatment of heavy quarkonium production is by far 
the most theoretically sound \cite{Bodwin:1994jh,Braaten:1996pv,Kramer:2001hh,Petrelli:1997ge}, 
and generally consistent with experimental data on inclusive production of $J/\psi$ and $\Upsilon$ of large transverse momentum, $p_T$, at the Tevatron and the LHC \cite{Ma:2010yw,Butenschoen:2010rq,Wang:2012is,Gong:2013qka}, it has not been able to explain fully the polarization of these heavy quarkonia produced at high-$p_T$ \cite{Affolder:2000nn,Abulencia:2007us,Acosta:2001gv,Abazov:2008aa}.  With a larger heavy quark mass, $m_Q$, it was expected that NRQCD factorization formalism should do a better job in describing the production of $\Upsilon$ so as its polarization.  However, recent data on polarization of $\Upsilon(1S,2S,3S)$ measured by CMS collaboration at the LHC \cite{Chatrchyan:2012woa} also shows inconsistency with full next-to-leading order (NLO) NRQCD calculation \cite{Gong:2013qka,Han:2014kxa}.  
In addition, global fits of data on $J/\psi$ production from various high energy collisions, including $e^+e^-$, lepton-hadron, and hadron-hadron collisions \cite{Butenschoen:2010rq,Butenschoen:2011yh} show some discrepancies in shape of momentum spectra between theory predictions and data \cite{Bodwin:2013nua}.  
For some production channels of the NRQCD calculations, the NLO corrections are orders larger than their corresponding leading order (LO) results, which raises questions as to whether yet higher order contributions can be neglected.  Motivated in part by these challenges to existing theory, new approaches based on QCD factorization \cite{Nayak:2005rt,Nayak:2006fm,Kang:2011zza,Kang:2011mg,Kang:2014tta} and soft-collinear effective theory \cite{Fleming:2012wy,Fleming:2013qu} have been proposed for the systematic study of heavy quarkonium production at collider energies.

In Ref.~\cite{Kang:2014tta}, we developed an extended QCD factorization formalism for heavy quarkonium production at large transverse momentum $p_T\gg m_H\gg \Lambda_{\rm QCD}$ in hadronic collisions 
(or at a large energy $E\gg m_H$ in $e^+e^-$ collisions) with heavy quarkonium mass $m_H$.  
The new QCD factorization formalism includes both collinear-factorized leading power (LP) and collinear-factorized
next-to-leading power (NLP) terms in the $1/p_T$ expansion of the production cross section, 
\begin{eqnarray}
\label{eq:pqcd_fac0}
E_P\frac{d\sigma_{A+B\to H+X}}{d^3P}(P)
&\approx &
\sum_{f} \int \frac{dz}{z^2} \, D_{f\to H}(z;m_Q) \, 
E_c\frac{d\hat{\sigma}_{A+B\to f(p_c)+X}}{d^3p_c}\left(p_c=\frac{1}{z}\, p\right) 
\nonumber\\
&+ & 
\sum_{[Q\bar{Q}(\kappa)]} \int \frac{dz}{z^2} \, du\, dv\
{\cal D}_{[Q\bar{Q}(\kappa)]\to H}(z,u,v;m_Q)
\\
&\ & 
\times 
E_c\frac{d\hat{\sigma}_{A+B\to [Q\bar{Q}(\kappa)](p_c)+X}}{d^3p_c}
\left(
P_Q=\frac{u}{z}\, p,P_{\bar{Q}}=\frac{\bar{u}}{z}\, p,
P'_Q=\frac{v}{z}\, p,P'_{\bar{Q}}=\frac{\bar{v}}{z}\, p\right),
\nonumber
\end{eqnarray}
where $p^\mu=P^\mu(m_H=0)$ is the massless part of the heavy quarkonium momentum in a frame 
in which the quarkonium moves along the $z$-axis,
and the renormalization scale $\mu$ and the factorization scale $\mu_F$ are suppressed. 
We will often refer to this result below as ``QCD factorization", to distinguish it from NRQCD factorization.
In this factorized expression, $\sum_f$ runs over all parton flavors $f=q,\bar{q},g$, 
including heavy flavors when $m_Q\ll p_T$, while $\sum_{[Q\bar{Q}(\kappa)]}$ indicates 
a sum over both spin and color states of heavy quark pairs $[Q\bar{Q}(\kappa)]$ 
where $\kappa=sI$ with $s=v,a,t$ for vector, axial vector and tensor spin states, and 
$I=1,8$ for singlet and octet color states, respectively. 
In Eq.~(\ref{eq:pqcd_fac0}), the variables $z$, $u$, and $v$, with $\bar{u}=1-u$ and $\bar{v}=1-v$ 
are light-cone momentum fractions; and $D_{f\to H}(z;m_Q)$ and 
${\cal D}_{[Q\bar{Q}(\kappa)]\to H}(z,u,v;m_Q)$ are single-parton and 
heavy quark-pair fragmentation functions (FFs), respectively \cite{Kang:2014tta}.
In the factorization formula in Eq.~(\ref{eq:pqcd_fac0}), we neglect all contributions 
involving twist-4 multi-parton correlation functions of colliding hadrons, as well as 
all terms at NLP involving FFs not from a heavy quark pair, because 
these contributions must create the heavy quark pair 
non-perturbatively.  We thus expect them to be suppressed by powers of heavy quark mass \cite{Kang:2014tta}, 
and could be suppressed further by reasons similar to those that lead to the OZI rule 
in evaluating the decay rates.

The QCD factorization formalism in Eq.~(\ref{eq:pqcd_fac0}) 
effectively organizes the contributions to heavy quarkonium production at large $p_T$ 
in terms of the characteristic time when an active heavy quark pair, 
which is necessary for a final state heavy quarkonuim, is produced.
The LP contribution to the production cross section is given by 
the hard partonic scattering to produce an active parton (quark, antiquark, or gluon) 
at a distance scale of ${\cal O}(1/p_T)$, convolved with a fragmentation function for this parton to 
evolve into a heavy quark pair that transmutes into a heavy quarkonium.  
At LP accuracy, the heavy quark pair is effectively produced at the distance scale of 
${\cal O}(1/2m_Q)$, a much longer distance compared to scales 
over which the active parton was initially produced.  
At NLP accuracy, the QCD factorization requires not only the factorized NLP term 
in Eq.~(\ref{eq:pqcd_fac0}), but also a new power suppressed contribution to  
the DGLAP evolution of heavy quarkonium FFs from a single active parton \cite{Kang:2014tta}. 
The factorized NLP term in Eq.~(\ref{eq:pqcd_fac0}) describes the production of 
the heavy quark pair directly at ${\cal O}(1/p_T)$ where the initial hard collision took place.  
The power suppressed contribution to the evolution of single parton FFs effectively sums 
up all leading logarithmic contributions to the production of the heavy quark pairs from 
the distance scale from ${\cal O}(1/\mu_F)$ to ${\cal O}(1/\mu_0)$ where
$\mu_F\sim p_T$ is the factorization scale and $\mu_0 \sim 2m_Q$ 
is the input scale at which the evolution of the FFs starts.
Having both LP and NLP contribution, the QCD factorization formalism in Eq.~(\ref{eq:pqcd_fac0})
effectively covers all leading contributions to the production of the heavy quark pair, 
which transmutes into an observed heavy quarkonium,  
no matter where and when the heavy quark pair was produced~\cite{Kang:2014tta}.  
If we keep only the factorized LP contribution to the cross section in
Eq.~(\ref{eq:pqcd_fac0}), we  include only the contribution to 
the heavy quarkonium production when the heavy quark pair is produced 
at the distance scale $\gtrsim {\cal O}(1/\mu_0)$.

The predictive power of the QCD factorization formalism in Eq.~(\ref{eq:pqcd_fac0}) 
relies on the universality of the FFs, and our ability to calculate the evolution kernels of 
these FFs, as well as the short-distance coefficient functions, perturbatively, to 
all orders in powers of $\alpha_s$.  In Ref.~\cite{Kang:2014tta},
we evaluated the mixing evolution kernels for one-parton to evolve into a heavy quark pair 
at ${\cal O}(\alpha_s^2)$, as well as evolution kernels for a heavy quark pair to evolve into
another heavy quark pair at ${\cal O}(\alpha_s)$.  
In this paper, we concentrate on the calculation of the short-distance coefficient functions 
of the QCD factorization formalism in Eq.~(\ref{eq:pqcd_fac0}).  
When $A$ and $B$ in Eq.~(\ref{eq:pqcd_fac0}) are hadrons, 
the cross section is found using following expressions, reflecting collinear factorization for the incoming hadrons,
\begin{eqnarray}
E_c\frac{d\hat{\sigma}_{A+B\to f(p_c)+X}}{d^3p_c}
&=& \sum_{a,b}
\int dx_a\, \phi_{A\to a}(x_a) \int dx_b\, \phi_{B\to b}(x_b) 
E_c\frac{d\hat{\sigma}_{a+b\to f(p_c)+X}}{d^3p_c} \, ,
\nonumber \\
E_c\frac{d\hat{\sigma}_{A+B\to [Q\bar{Q}(\kappa)](p_c)+X}}{d^3p_c}
&=& \sum_{a,b}
\int dx_a\, \phi_{A\to a}(x_a) \int dx_b\, \phi_{B\to b}(x_b) 
E_c\frac{d\hat{\sigma}_{a+b\to [Q\bar{Q}(\kappa)](p_c)+X}}{d^3p_c}
\, ,
\label{eq:pqcd_fac_h}
\end{eqnarray}
where $a,b$ represent active parton flavors, running over quarks, antiquarks and gluons,
and $\phi_{A\to a}(x)$ and $\phi_{B\to b}(x)$ are the parton distribution functions (PDFs) 
of hadron $A$ and $B$, respectively, with factorization scale dependence suppressed.
The short-distance coefficient functions for producing a single parton at the LP, 
$\hat{\sigma}_{a+b\to f(p_c)+X}$ in Eq.~(\ref{eq:pqcd_fac0}), 
are the same as the perturbative coefficient functions for producing a light hadron, 
such as pion, and are available for both the LO 
and NLO in powers of $\alpha_s$ in the literature \cite{Aversa:1988vb}. 
This is because the factorized short distance coefficient functions 
are not sensitive to the details of the hadron produced in the final state,
but only the properties of the fragmenting parton.
In the next section, we introduce the method to calculate the short-distance hard parts 
at NLP, $\hat{\sigma}_{a+b\to [Q\bar{Q}(\kappa)](p_c)+X}$ in Eq.~(\ref{eq:pqcd_fac_h}), 
and present the detailed calculations of the ${\cal O}(\alpha_s^3)$ coefficient functions 
for all relevant spin-color states of a heavy quark pair produced from the scattering of 
a light quark and antiquark.  The complete results of short-distance hard parts 
for all parton-parton scattering channels at ${\cal O}(\alpha_s^3)$ are given 
in Appendix~\ref{app:hardparts}.

With the perturbatively calculated short-distance hard parts in this paper, 
and the evolution kernels derived in Ref.~\cite{Kang:2014tta}, the predictive power of 
QCD factorization formalism in Eq.~(\ref{eq:pqcd_fac0}) still requires our knowledge 
of the universal FFs at an input scale $\mu_0$.  
Since both the short-distance hard parts and evolution kernels are perturbative and 
the same for hadronic production of all heavy quarkonium states, 
it is these input FFs that carry the information on 
the characteristics of the individual heavy quarkonia.
The universal input FFs are non-perturbative, and in principle, 
should be extracted from fitting experimental data, 
like the FFs for inclusive light hadron production.
However, with the NLP contributions, we will need many more FFs 
(single parton plus heavy quark pair FFs) for each heavy quarkonium
state produced.  Extracting all these FFs from inclusive cross sections of heavy
quarkonium production would not be an easy task.

Unlike the FFs to a light hadron, heavy quarkonium FFs at the input
scale $\mu_0$ have effectively one intrinsic hard scale 
-- heavy quark mass $m_Q \sim {\cal O}(\mu_0)$, 
which is sufficiently separated from the momentum scale 
for the binding of heavy quarkonium, $m_Q v$ with  
the heavy quark relative velocity in the pair's rest frame, $v \ll 1$.  
The physics between the $m_Q$ and $\mu_0$ could be perturbatively calculable. 
It was proposed in Ref.\ \cite{Kang:2011mg}, as a conjecture or a model, to use NRQCD factorization 
to calculate these input FFs by expressing all of them in terms of 
perturbatively calculated coefficients and a few local 
NRQCD matrix elements, organized in powers of $v$.
Since the input momentum scale $\mu_0 \sim \mu_\Lambda$, the NRQCD factorization scale
$\sim {\cal O}(m_Q)$, the perturbatively calculated coefficient functions should be free of the large
logarithms and the power enhancement that were seen in the NLO NRQCD 
coefficient functions for heavy quarkonium production at large $p_T$ at collider energies 
\cite{Ma:2010yw,Butenschoen:2010rq,Wang:2012is,Gong:2013qka}.  
In Sec.~\ref{sec:nrqcd-ffs}, we review the procedure to calculate the heavy quarkonium FFs 
at the input scale $\mu_0 \gtrsim 2m_Q$ in terms of NRQCD factorization.  

Although there is no formal proof that ensures that NRQCD factorization works for 
evaluating these universal input FFs perturbatively to all orders in $\alpha_s$
and all powers in $v$-expansion, it has been demonstrated that such NRQCD factorization 
should work up to two-loop radiative corrections \cite{Nayak:2005rt,Nayak:2006fm}. 
Explicit perturbative calculations in Refs.~\cite{Ma:2013yla,Ma:2014eja} show that
such factorization is indeed possible for up to $v^4$ in the velocity expansion 
since all calculated perturbative coefficient functions are infrared safe 
for LP single-parton FFs at ${\cal O}(\alpha_s^2)$, 
as well as for NLP heavy quark-pair FFs at ${\cal O}(\alpha_s)$.  
Such perturbatively calculated input FFs in NRQCD factorization should provide 
a good starting point to estimate or determine these much needed universal 
but nonperturbative functions for heavy quarkonium production.

If the NRQCD factorization for calculating the input FFs is valid, the collinear factorization
formalism in Eq.~(\ref{eq:pqcd_fac0}) is effectively a reorganization of the perturbatively 
calculated cross section by NRQCD factorization, with resummation of large fragmentation
logarithms.  It also provides a justification of NRQCD factorization applied to heavy
quarkonium production at large transverse momentum, at least for the first and second 
power terms in the $1/p_T$ expansion.  In Sec.~\ref{sec:nrqcd-ffs}, 
we discuss the connection between the QCD factorization formalism in 
Eq.~(\ref{eq:pqcd_fac0}) and the NRQCD factorization approach to 
heavy quarkonium production \cite{Bodwin:1994jh}.  
With a proper matching, we introduce an expanded factorization formalism 
which could smoothly connect the QCD factorization in Eq.~(\ref{eq:pqcd_fac0}) 
for $p_T\gg \mu_0$ to the fixed-order calculation in NRQCD factorization for 
$p_T\gtrsim \mu_0\gtrsim 2m_Q$, including heavy quark mass effects.  

In subsection~\ref{subsec:singlet}, we provide an explicit example to demonstrate that 
when $p_T\gg m_H$, the QCD factorization formalism in Eq.~(\ref{eq:pqcd_fac0}) 
catches all leading contributions to the heavy quarkonium production.  We show that 
the extremely challenging calculation of the complete NLO contributions to 
the production of a color singlet, spin-1 heavy quark pair in hadronic collisions 
can be effectively reproduced by the much simpler LO perturbative QCD calculation 
of the hard parts to produce a color-octet collinear and massless heavy quark pair,
convolved with equally simple LO fragmentation functions for the perturbatively 
produced pair to fragment into the color-singlet, spin-1 heavy quark pair, 
calculated in NRQCD factorization.  The combination of the two LO calculations
 reproduces more than 95\% of the full NLO contribution when $p_T$ is 
only a few times of the heavy quark mass.  The same conclusion is also
true for other production channels in NRQCD calculations \cite{Ma:2014svb}.  

In Sec.~\ref{sec:pol}, we discuss how to evaluate heavy quarkonium polarization 
in the QCD factorization approach.  Since both short-distance partonic hard parts and 
evolution kernels of heavy quarkonium FFs are perturbative, and not sensitive to 
the long-distance details of the individual heavy quarkonium produced, the heavy quarkonium 
polarization should be completely determined by the heavy quarkonium FFs 
at the input factorization scale, $\mu_0$.  With $\mu_0\gtrsim m_Q \gg m_Q v$, 
it is very reasonable to apply the same NRQCD factorization conjecture 
for calculating the unpolarized heavy quarkonium FFs at scale $\mu_0$ to the 
calculation of polarized heavy quarkonium FFs at the same input scale.  In this section,
we present the projection operators, within the NRQCD factorization approach, 
for the calculation of polarized heavy quarkonium FFs with the produced heavy quarkonium 
in either a transverse or a longitudinal polarization state.  
As an example, we present our ${\cal O}(\alpha_s)$ calculation of 
polarized heavy quarkonium FFs via a color singlet $\CScSa$ heavy quark pair in NRQCD.  
A complete calculation of polarized heavy quarkonium FFs in NRQCD 
for all partonic channels is now available 
\cite{Ma:2013yla,Ma:2014eja,Zhang:thesis}.
With the perturbatively calculated polarized heavy quarkonium FFs, 
we demonstrate explicitly that the combination of QCD factorization 
for heavy quarkonium production in Eq.~(\ref{eq:pqcd_fac0}) and 
NRQCD factorization for the heavy quarkonium FFs can not only 
reproduce the NLO Color Singlet Model (CSM) calculation for the production rate, 
but also the polarization of the produced heavy quarkonia.
Clearly, a fuller understanding of heavy quarkonium production and its polarization
requires a new global analysis of all heavy quarkonium production data in terms
of QCD factorization formalism and the new set of evolution equations for
heavy quarkonium FFs \cite{Ma:2014svb}.  Finally, our conclusions are summarized 
in Sec.~\ref{sec:summary}.

\newpage
\section{Production cross section and partonic hard parts}
\label{sec:hardparts}

In this section, we introduce a systematic method to calculate all partonic 
hard parts of the collinear-factorized NLP terms, $d\hat{\sigma}_{ab\to[Q\bar{Q}(\kappa)](p_c)}$,  
in Eq.~(\ref{eq:pqcd_fac_h}) perturbatively.  We provide a detailed derivation of the hard parts 
for producing a heavy quark pair in various spin-color states from the scattering of 
a quark and an antiquark at ${\cal O}(\alpha_s^3)$, and present our full results for
all other partonic scattering channels, including quark-gluon and gluon-gluon scattering 
channels in the Appendix~\ref{app:hardparts}.

\subsection{The formalism}
\label{subsec:formula}

As a consequence of QCD factorization, all factorized partonic hard parts for heavy quarkonium production in Eq.~(\ref{eq:pqcd_fac_h}) are uniquely determined perturbatively by the factorization formalism and the definition of fragmentation functions (and parton distribution functions in the case of hadronic collisions).  
Like the LP hard parts, the NLP factorized partonic hard parts, 
$d\hat{\sigma}_{ab\to[Q\bar{Q}(\kappa)](p_c)}$ in Eq.~(\ref{eq:pqcd_fac_h}), 
are insensitive to the long-distance details of the colliding hadrons and 
the produced heavy quarkonium.  The factorization formalism in Eq.~(\ref{eq:pqcd_fac0})
is also valid when the colliding hadrons, $A$ and $B$, are replaced 
by two asymptotic colliding parton states of flavor $a$ and $b$, respectively, 
and the produced heavy quarkonium, $H$, is replaced by an asymptotic state 
of a heavy quark pair with momentum $p$ and spin-color state $[Q\bar{Q}(\kappa)]$.
Together, the partonic analogs of Eqs.\ (\ref{eq:pqcd_fac0}) and (\ref{eq:pqcd_fac_h}) 
can  be expressed symbolically as,
\begin{eqnarray}
\label{eq:pqcd_fac}
d\sigma_{a+b\to [Q\bar{Q}(\kappa)](p)}
&\approx &
\sum_{i,j,f} \phi_{a\to i} \otimes \phi_{b\to j} \otimes 
d\hat{\sigma}_{i+j\to f(p_c)} \otimes D_{f\to [Q\bar{Q}(\kappa)]}
\nonumber\\
&+ & 
\sum_{i,j,[Q\bar{Q}(\kappa')]} 
\phi_{a\to i} \otimes \phi_{b\to j} \otimes 
d\hat{\sigma}_{i+j\to [Q\bar{Q}(\kappa')](p_c)} \otimes
{\cal D}_{[Q\bar{Q}(\kappa')]\to [Q\bar{Q}(\kappa)]}\, ,
\end{eqnarray}
where $i,j,f$ represent the factorized active parton flavors, including $q, \bar{q}, g$, and $Q$, 
$[Q\bar{Q}(\kappa')]$ represents a heavy quark pair of spin-color state $\kappa'$, and
$\otimes$ represents the convolution over partonic momentum fractions as
shown in Eqs.~(\ref{eq:pqcd_fac0}) and (\ref{eq:pqcd_fac_h}).
Unlike the cross section in Eq.~(\ref{eq:pqcd_fac0}), both the partonic cross section on the
left-hand-side (LHS), and the PDFs of a parton and FFs of a partonic state 
on the right-hand-side (RHS) of Eq.~(\ref{eq:pqcd_fac}) can be  calculated perturbatively
in terms of Feynman diagrams with proper regularizations.  Most importantly, the short-distance 
partonic hard parts in Eq.~(\ref{eq:pqcd_fac}) are the same as those in Eq.~(\ref{eq:pqcd_fac_h}).  

The fact that the cross section is factorizable ensures that the perturbatively calculated partonic
cross sections on the LHS and PDFs and FFs on the RHS of Eq.~(\ref{eq:pqcd_fac}) 
are all free of infrared (IR) divergence, 
while the ultraviolet (UV) divergences are taken care of by the renormalization, and 
all CO divergences are process-independent and cancelled perturbatively
order-by-order in powers of $\alpha_s$ between the LHS and the RHS to leave the partonic hard parts
free of any divergences.  To derive the partonic hard parts in Eq.~(\ref{eq:pqcd_fac_h}), which are the
same as those in Eq.~(\ref{eq:pqcd_fac}), we expand both sides of Eq.~(\ref{eq:pqcd_fac})
order-by-order in powers of $\alpha_s$, and then, extract all partonic hard parts perturbatively
by calculating the corresponding partonic cross section in the LHS, and the PDFs and FFs of partons 
on the RHS.  

\begin{figure}[!htp]
\noindent
\includegraphics[width=0.25\columnwidth]{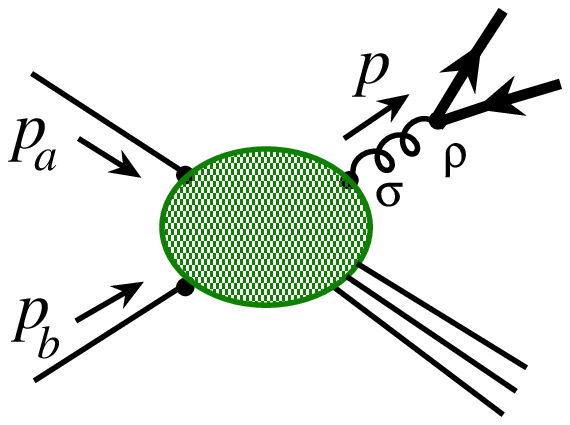}
\caption{Sample partonic scattering amplitude for $a(p_a)+b(p_b)\to [Q\bar{Q}(\kappa)](p)+X$ 
that includes a leading power contribution to the production rate of a heavy quark pair. }
\label{fig:ab2QQx}
\end{figure}

To evaluate the hard parts at the first non-trivial order in hadronic collisions, 
we expand both sides of Eq.~(\ref{eq:pqcd_fac}) to ${\cal O}(\alpha_s^3)$,
\begin{eqnarray}
\label{eq:pqcd_fac3}
d\sigma^{(3)}_{a+b\to [Q\bar{Q}(\kappa)](p)}
&\approx &
\sum_{i,j,f} \phi^{(0)}_{a\to i} \otimes \phi^{(0)}_{b\to j} \otimes 
d\hat{\sigma}^{(2)}_{i+j\to f(p_c)} \otimes D^{(1)}_{f\to [Q\bar{Q}(\kappa)]}
\nonumber\\
&+ & 
\sum_{i,j,[Q\bar{Q}(\kappa')]} 
\phi^{(0)}_{a\to i} \otimes \phi^{(0)}_{b\to j} \otimes 
d\hat{\sigma}^{(3)}_{i+j\to [Q\bar{Q}(\kappa')](p_c)} \otimes
{\cal D}^{(0)}_{[Q\bar{Q}(\kappa')]\to [Q\bar{Q}(\kappa)]}\, ,
\end{eqnarray}
where the superscript $(m)$ with $m=0,1,2,3$ indicates the power of $\alpha_s$ of the corresponding quantity.  Since the zeroth order parton PDFs and FFs are given by the $\delta$-functions that fix the corresponding convolutions, the short-distance hard parts 
for all possible channels of partonic scattering between parton flavors $a$ and $b$ are given by
\begin{eqnarray}
\label{eq:ab2QQb}
d\hat{\sigma}^{(3)}_{a+b\to [Q\bar{Q}(\kappa)](p)}
= 
d\sigma^{(3)}_{a+b\to [Q\bar{Q}(\kappa)](p)}
-
d\hat{\sigma}^{(2)}_{a+b\to g(p_c)} \otimes D^{(1)}_{g\to [Q\bar{Q}(\kappa)](p)} \, ,
\end{eqnarray}
where the first term on the RHS, $d\sigma^{(3)}_{a+b\to [Q\bar{Q}(\kappa)](p)}$, 
is the perturbative cross section for two partons of flavors $a$ and $b$ to 
produce a heavy quark pair of momentum $p$ in a spin-color quantum state $[Q\bar{Q}(\kappa)]$ 
at order of $\alpha_s^3$, which covers all 
LO partonic processes to produce a heavy quark pair at a large transverse momentum.  
Since we neglect the heavy quark mass when $p_T\gg m_Q$, this partonic scattering 
amplitude, like the one in Fig.~\ref{fig:ab2QQx}, can have a perturbative
divergence caused by the mass singularity of the gluon propagator, 
when its invariant mass goes on-shell, $p^2\to 0$, which leads to a divergent 
partonic cross section, $d\sigma^{(3)}_{a+b\to [Q\bar{Q}(\kappa)](p)}$.  
This perturbatively divergent contribution is in fact a LP contribution, and
is already included in the LP fragmentation contribution, 
the first term on the RHS of Eq.~(\ref{eq:pqcd_fac0}).  Therefore, it 
should be systematically removed when we calculate the short-distance
partonic hard parts of the NLP contribution to avoid double counting.
In Eq.~(\ref{eq:ab2QQb}), the second term 
is a natural result of the QCD factorization formalism.  Its role 
is to remove all possible LP contributions from the first term, and it 
can be referred as a subtraction term for removing the mass singularity 
or the LP contribution.

Similarly, by expanding the factorized formalism for the partonic scattering cross section 
in Eq.~(\ref{eq:pqcd_fac}) to  order  $\alpha_s^4$, we derive the factorization formula 
for calculating the NLO short-distance partonic hard parts of the NLP contribution as,
\begin{eqnarray}
d\hat{\sigma}^{(4)}_{a+b\to [Q\bar{Q}(\kappa)](p)}
&=& 
d\sigma^{(4)}_{a+b\to [Q\bar{Q}(\kappa)](p)}
-
d\hat{\sigma}^{(3)}_{a+b\to g(p_c)} \otimes D^{(1)}_{g\to [Q\bar{Q}(\kappa)](p)} 
\nonumber\\
&\ &
- \sum_{f}
d\hat{\sigma}^{(2)}_{a+b\to f(p_c)} \otimes D^{(2)}_{f\to [Q\bar{Q}(\kappa)](p)} 
\nonumber\\
&\ &
- \sum_{i}
\phi_{a\to i}^{(1)}\otimes
d\hat{\sigma}^{(2)}_{i+b\to g(p_c)} \otimes D^{(1)}_{g\to [Q\bar{Q}(\kappa)](p)} 
\nonumber\\
&\ &
- \sum_{j}
\phi_{b\to j}^{(1)}\otimes
d\hat{\sigma}^{(2)}_{a+j\to g(p_c)} \otimes D^{(1)}_{g\to [Q\bar{Q}(\kappa)](p)} 
\nonumber\\
&\ &
- \sum_{i}
\phi_{a\to i}^{(1)}\otimes
d\hat{\sigma}^{(3)}_{i+b\to [Q\bar{Q}(\kappa)](p)} 
- \sum_{j}
\phi_{b\to j}^{(1)}\otimes
d\hat{\sigma}^{(3)}_{a+j\to [Q\bar{Q}(\kappa)](p)} \, ,
\label{eq:ab2QQbNLO}
\end{eqnarray}
where the sum of $i,j,f$ runs over all parton flavors,  and 
all lower order short-distance partonic hard parts
are well-defined and calculable.  For example, the 
$d\hat{\sigma}^{(3)}_{a+b\to g(p_c)}$ are given by Eq.~(\ref{eq:ab2QQb}), and 
$d\hat{\sigma}^{(2)}_{a+b\to f(p_c)}$ are the lowest order partonic cross sections 
given by lowest order $2\to 2$ partonic scattering amplitudes, and are finite.
In Eq.~(\ref{eq:ab2QQbNLO}), the subtraction term in the first line 
plays the same role as that of the subtraction term in Eq.~(\ref{eq:ab2QQb}), 
the subtraction term in the second line is to 
remove the power collinear divergence of the partonic cross section, 
$d\sigma^{(4)}_{a+b\to [Q\bar{Q}(\kappa)](p)}$, which has been included 
in the evolution of the single parton FFs via the mixing kernels from 
a single fragmenting parton to a heavy quark pair \cite{Kang:2014tta}. 
The four more subtraction terms in the last three lines of Eq.~(\ref{eq:ab2QQbNLO}) 
are needed to remove the logarithmic collinear contributions that 
have been included in the evolution of initial-state parton distribution functions (PDFs).  
The factorization formula in Eq.~(\ref{eq:ab2QQbNLO}) 
can be adapted for calculating the NLO contribution of the power
corrections in other scattering processes, for example, 
we only need the first two lines for high energy heavy quarkonium production 
in $e^+e^-$ collisions.

\begin{figure}[!htp]
\noindent
\includegraphics[width=0.17\columnwidth]{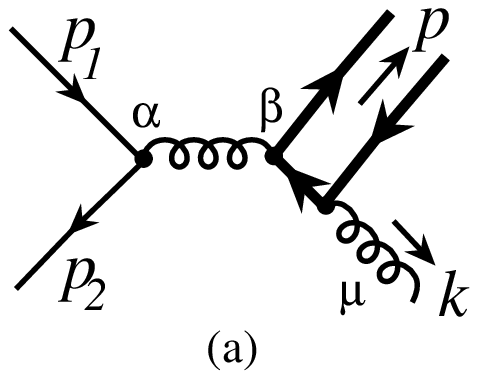}
\hskip 0.02\columnwidth
\includegraphics[width=0.17\columnwidth]{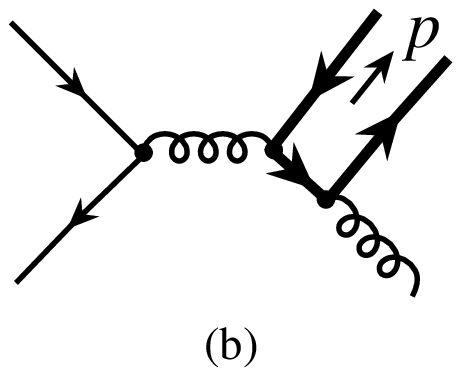}
\hskip 0.02\columnwidth
\includegraphics[width=0.17\columnwidth]{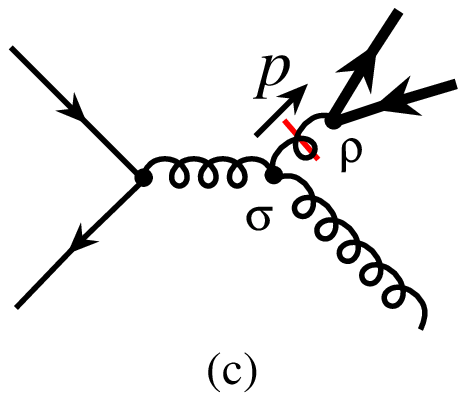}
\hskip 0.02\columnwidth
\includegraphics[width=0.17\columnwidth]{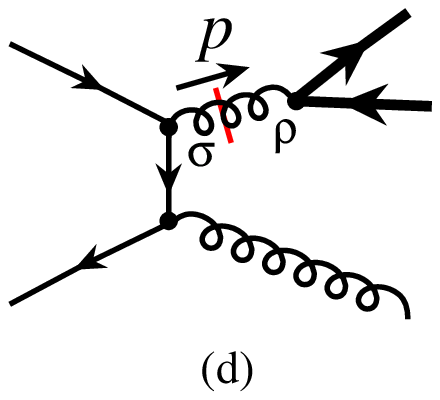}
\hskip 0.02\columnwidth
\includegraphics[width=0.17\columnwidth]{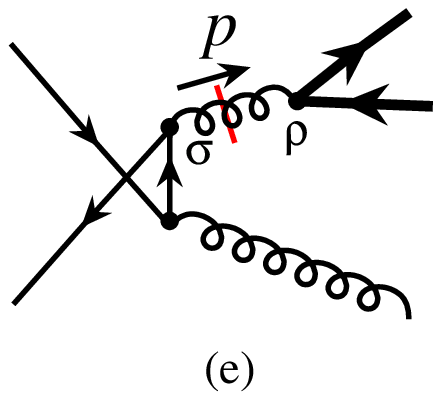}
\caption{Leading order Feynman diagrams for the $q\bar{q}\to Q\bar{Q}g$ subprocess 
at ${\cal O}(\alpha_s^3)$.}
\label{fig:qq2QQg}
\end{figure}

In the remainder of this section, we provide the detailed derivation of the first non-trivial 
short-distance hard parts for a quark and an antiquark to produce a heavy quark pair 
in all possible spin and color states.   Because the heavy quark mass is set to zero in 
the hard parts, we only need to consider the production of a heavy quark pair 
in axial-vector and vector spin states at the order of $\alpha_s^3$.
That is, we calculate the partonic hard parts by using Eq.~(\ref{eq:ab2QQb}) 
with $a=q(p_1)$ and $b=\bar{q}(p_2)$ of momentum $p_1$ and $p_2$, respectively,  
and
\begin{eqnarray}
\label{eq:qqb2QQb}
d\hat{\sigma}^{(3)}_{q(p_1)+\bar{q}(p_2)\to [Q\bar{Q}(\kappa)](p)}
= 
d\sigma^{(3)}_{q(p_1)+\bar{q}(p_2)\to [Q\bar{Q}(\kappa)](p)}
-
d\hat{\sigma}^{(2)}_{q(p_1)+\bar{q}(p_2)\to g(p_c)} 
\otimes D^{(1)}_{g(p_c)\to [Q\bar{Q}(\kappa)](p)} 
\end{eqnarray}
where the produced pair $[Q\bar{Q}(\kappa)]$ can be in an axial-vector or a vector spin state 
while in either singlet or octet color state. 

\begin{figure}[!htp]
\noindent
\includegraphics[width=0.17\columnwidth]{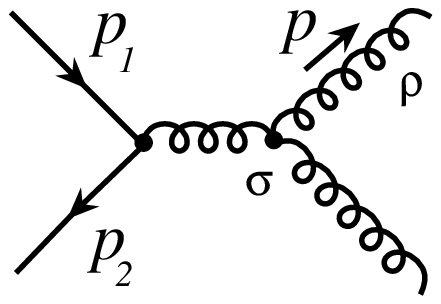}
\hskip 0.04\columnwidth
\includegraphics[width=0.17\columnwidth]{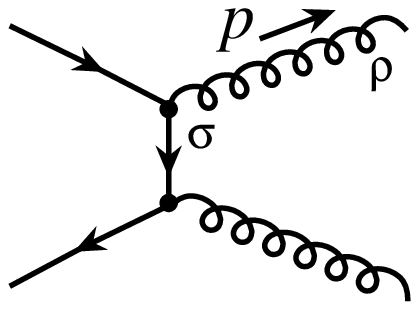}
\hskip 0.04\columnwidth
\includegraphics[width=0.17\columnwidth]{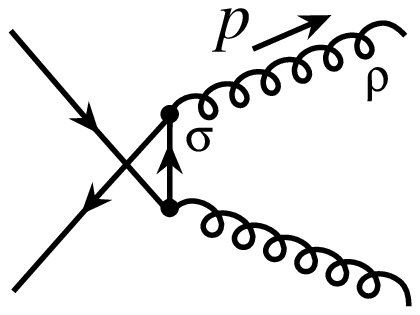}
\caption{Leading order Feynman diagrams for $q\bar{q}\to gg$ subprocess
at ${\cal O}(\alpha_s^2)$.}
\label{fig:qqb2gg}
\end{figure}

\begin{figure}[!htp]
\noindent
\includegraphics[width=0.28\columnwidth]{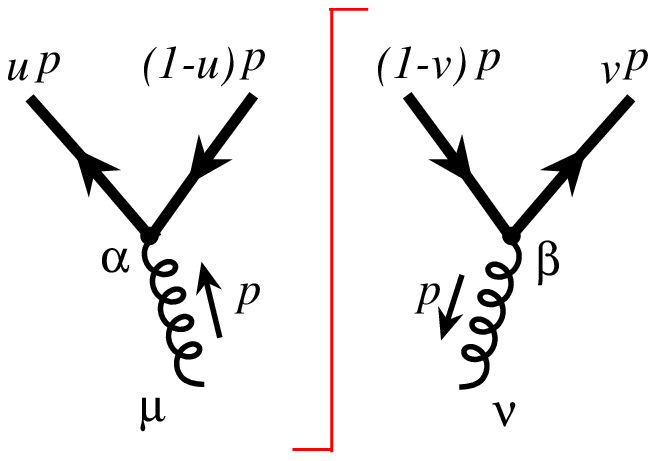}
\caption{Lowest order Feynman diagram ($\alpha_s$) for a gluon to fragment into a heavy quark pair.}
\label{fig:g2QQbLO}
\end{figure}

At the order of $\alpha_s^3$, the scattering amplitude 
for $d\sigma^{(3)}_{q(p_1)+\bar{q}(p_2)\to [Q\bar{Q}(\kappa)](p)}$
is given by the Feynman diagrams in Fig.~\ref{fig:qq2QQg}.
The diagrams in Fig.~\ref{fig:qq2QQg}(c), (d), and (e) all have divergent contributions 
caused by the same mass singularity when the momentum of the gluon 
with a short bar goes on-shell, $p^2\to 0$.  
As discussed above, any such perturbatively-divergent LP contribution 
should be removed by a subtraction term, the second term 
on the RHS of Eq.~(\ref{eq:qqb2QQb}).
The $d\hat{\sigma}^{(2)}_{q(p_1)+\bar{q}(p_2)\to g(p_c)}$ of the subtraction term
is the lowest order cross section for partonic process, 
$q(p_1)+\bar{q}(p_2) \to g(p) + g$, 
given by the Feynman diagrams in Fig.~\ref{fig:qqb2gg}.
At order $\alpha_s^2$, the partonic cross section 
$d\hat{\sigma}^{(2)}_{q(p_1)+\bar{q}(p_1)\to g(p_c)}$ is perturbatively finite.
The function $D^{(1)}_{g(p_c)\to [Q\bar{Q}(\kappa)](p)}$ of the subtraction term 
in Eq.~(\ref{eq:qqb2QQb})
is the lowest order fragmentation function for a gluon to a heavy quark pair.
At order $\alpha_s$, it is given by the Feynman diagram in Fig.~\ref{fig:g2QQbLO}, 
which is in cut diagram notation, where the amplitude and complex conjugate 
are combined into a forward scattering diagram and 
the final state is identified by a vertical line.
From the definition of the gluon fragmentation function, 
two gluon lines in Fig.~\ref{fig:g2QQbLO} are contracted by the cut vertex \cite{Kang:2014tta},
\begin{eqnarray}
{\cal V}_g(z) = \int \frac{d^4p_c}{(2\pi)^4}\ z^2 \delta\left(z-\frac{p\cdot \hat{n}}{p_c\cdot \hat{n}}\right) 
\left[\frac{1}{N_c^2-1}\sum_{a=1}^{N_c^2-1}\ \delta_{a'a}\
\left( \frac{1}{2} \widetilde{d}_{\hat{n}}^{\mu\nu}(p_c) \right)\right]
\label{eq:cv_g}
\end{eqnarray}
where the four-vector $\hat{n}^\mu$ with $\hat{n}^2=0$ is an auxiliary vector 
conjugate to the observed hadron momentum $p^\mu$, introduced to help define 
the fragmenting gluon's light-cone momentum fraction, as well as its
two transverse polarization states (or ``physical'' polarization states).
In Eq.~(\ref{eq:cv_g}), the $a$ and $a'$ are color indices of the fragmenting gluon 
in the amplitude and its complex conjugate, respectively, and 
\begin{equation}
\widetilde{d}_{\hat{n}}^{\mu\nu}(p_c)
=-g^{\mu\nu} + \frac{p_c^\mu \hat{n}^\nu + \hat{n}^\mu p_c^\nu}{p_c\cdot \hat{n}} 
-\frac{p_c^2}{(p_c\cdot \hat{n})^2}\, \hat{n}^\mu \hat{n}^\nu\, ,
\label{eq:cv_g0}
\end{equation}
with $\widetilde{d}_{\hat{n}}^{\mu\nu}(p_c)\, p_{c\mu}  = 
\widetilde{d}_{\hat{n}}^{\mu\nu}(p_c)\, \hat{n}_\mu = 0$.
In a frame where the hadron is moving along the $+z$-axis, $p^\mu=(p^+,0^-,0_\perp)$ 
with hadron mass neglected, we can normalize the auxiliary vector $\hat{n}^\mu$ as 
$\hat{n}^\mu = (0^+,1^-,0_\perp)$, since the cut vertex, 
${\cal V}_g(z)$, is invariant when we rescale the vector $\hat{n}^\mu$.  
At the lowest order, the fragmenting gluon momentum $p_c$ above is effectively 
equal to the momentum of the heavy quark pair $p$ in Fig.~\ref{fig:g2QQbLO}.
Consequently, the LO perturbative gluon FF, $D^{(1)}_{g(p_c)\to [Q\bar{Q}(\kappa)](p)}$, 
is divergent as the gluon momentum goes on-shell, $p_c^2\to p^2\to 0$.
It is clear from above discussion that the second term in Eq.~(\ref{eq:qqb2QQb})
matches precisely the structure of the divergent piece of the partonic cross section 
$d\sigma^{(3)}_{q(p_1)+\bar{q}(p_2)\to [Q\bar{Q}(\kappa)](p)}$ to remove its mass singularity
when $p^2\to 0$, and to leave the sum of these two terms in Eq.~(\ref{eq:qqb2QQb}) 
infrared safe (IRS) and perturbative.  
This subtraction also avoids double counting of the LP contribution, 
as required by QCD factorization.

The cancelation of the divergence between the first and the second terms 
in Eq.~(\ref{eq:qqb2QQb}) is exact at the phase space point 
where the gluon with a short bar in Figs.~\ref{fig:qq2QQg}e and f
is on the mass-shell, with a physical polarization.  
It was shown in Ref.~\cite{Kang:2014tta}, as part of the calculation of the 
evolution kernels for a single parton to evolve into a heavy quark pair, that
the net effect of the second term in Eq.~(\ref{eq:qqb2QQb}) is to replace each
gluon propagators with the short bar in Fig.~\ref{fig:qq2QQg} by a contact term, 
found by rewriting the propagator as 
\begin{eqnarray}
G^{\rho\sigma}(p) 
&=& \frac{i}{p^2+i\varepsilon} \left[
-g^{\rho\sigma} + \frac{p^\rho \hat{n}^\sigma + p^\sigma \hat{n}^\rho}{p\cdot\hat{n}}
- \frac{p^2}{(p\cdot\hat{n})^2} \hat{n}^\rho \hat{n}^\sigma \right]
+
\frac{i}{p^2+i\varepsilon} \left[\frac{p^2}{(p\cdot\hat{n})^2} \hat{n}^\rho \hat{n}^\sigma \right]\, ,
\nonumber\\
\label{eq:gluonp}
\end{eqnarray}
and keeping only the final, unphysical term,
\begin{eqnarray}
G^{\rho\sigma}(p) 
&\to &
\frac{i\, p^2}{p^2 + i\varepsilon}\left[ \frac{\hat{n}^\rho \hat{n}^\sigma }{(p\cdot \hat{n})^2}\right]
\equiv
G^{\rho\sigma}_c(p)
\label{eq:gluonsp}\, ,
\end{eqnarray}
where (\ref{eq:gluonsp}) shows the result of subtracting the contribution of these diagrams to gluon fragmentation, which
is already included in the LP term.

As in Eq.\ (\ref{eq:cv_g0}), the first term in the right-hand side of Eq.\  (\ref{eq:gluonp}) vanishes when it is contracted 
by either $p_\rho$ or $\hat{n}_\rho$, defining the contribution from the gluon's
two transverse polarization states.    These are precisely the terms cancelled by the subtraction, which takes the
physical polarizations into account.
In Eq.~(\ref{eq:gluonp}), the second term in the RHS, is the  contact term, 
which is finite when the gluon line goes on-shell, $p^2\to 0$.
With this term included, the  NLP contribution to the scattering amplitude 
in Fig.~\ref{fig:qq2QQg} is color gauge invariant \cite{Qiu:1988dn}. 
Then, the effect of the subtraction  is to specify 
a prespcription for calculating the hard parts at this order
following Eq.~(\ref{eq:qqb2QQb}), which can be represented as \cite{Kang:2014tta}
\begin{eqnarray}
\label{eq:qqb2QQbgc}
d\hat{\sigma}^{(3)}_{q+\bar{q}\to [Q\bar{Q}(\kappa)](p)}(p_1,p_2,p)
= 
d\sigma^{(3)}_{q+\bar{q}\to [Q\bar{Q}(\kappa)](p)}(p_1,p_2,p)_c\, ,
\end{eqnarray}
where the subscript ``$c$'' indicates that the gluon propagators 
with a short bar in Fig.~\ref{fig:qq2QQg} are replaced by corresponding contact terms.    
These diagrams, in Figs.~\ref{fig:qq2QQg}$(c), (d)$ and $(e)$,  
with their perturbatively divergent leading power contributions removed, 
are necessary for the gauge invariance of the NLP contribution.  
In addition to quark-antiquark scattering, the expression in Eq.~(\ref{eq:qqb2QQbgc}) 
is also valid for scattering of two partons of any flavors $a$ and $b$ at this order.

In general, the removal of the LP contributions to the partonic scattering cross sections, 
or more specifically, the cancelation of divergences when $p^2\to 0$ between the first and 
the second terms in Eq.~(\ref{eq:ab2QQb}), or the terms in Eq.~(\ref{eq:ab2QQbNLO}), 
can be handled by introducing a regulator for the divergence of each term first, 
calculating all terms individually, and then
removing the regulator after all terms are combined and divergent terms are cancelled.  
Such a general approach  for calculating partonic hard parts 
beyond the LO contribution derived here could be made algorithmic.

Having identified the contact-term prescription, Eq.~(\ref{eq:qqb2QQbgc}), 
it is now straightforward to calculate NLP partonic hard parts
for all partonic scattering channels at ${\cal O}(\alpha_s^3)$, once we specify 
the projection operators for the  spin-color states of 
the produced heavy quark pair, $[Q\bar{Q}(\kappa)]$.  
The perturbatively produced collinear heavy quark pair should have four spin states, 
$s=v,a,t$ for vector, axial vector, and two tensor states, respectively, 
and nine color states, $I=1,8$ for singlet and octet color states, respectively.  
The corresponding projection operators have been defined in Ref.\ \cite{Kang:2014tta},
\begin{eqnarray}
\widetilde{\cal P}^{(v)}(p)_{ji,kl}
&=& 
\left(\gamma\cdot p\right)_{ji}\,
\left(\gamma\cdot p \right)_{kl}
\, ,
\nonumber\\
\widetilde{\cal P}^{(a)}(p)_{ji,kl}
&=& 
\left(\gamma\cdot p\, \gamma_5 \right)_{ji}\,
\left(\gamma\cdot p\, \gamma_5 \right)_{kl}
\, ,
\nonumber\\
\widetilde{\cal P}^{(t)}(p)_{ji,kl}
&=& 
\sum_{\alpha=1,2}\left(\gamma\cdot p\gamma_\perp^\alpha \right)_{ji}\,
\left(\gamma\cdot p\gamma_\perp^\alpha \right)_{kl}
\, ,
\label{eq:spin-pj}
\end{eqnarray}
for spin states of the produced heavy quark pair, and
\begin{eqnarray}
\widetilde{\cal C}_{ba,dc}^{[1]} &= &
\left[\frac{\delta_{ba}}{\sqrt{N_c}} \right] \left[ \frac{\delta_{dc}}{\sqrt{N_c}} \right] \, ,
\nonumber\\
\widetilde{\cal C}_{ba,dc}^{[8]} &= &
\sum_{A} \left[\sqrt{2} \left(t^A\right)_{ba}\right]\, 
               \left[\sqrt{2} \left(t^A\right)_{dc}\right]\, .
\label{eq:color-pj}
\end{eqnarray}
for the color states of the same pair.  
In Eq.~(\ref{eq:spin-pj}), the spin projection operators are independent of the momentum fractions 
of the produced heavy quark and antiquark, and the subscripts, $ji$ and $kl$ represent 
the spinor indices of the heavy quark pair in the scattering amplitude and  its complex 
conjugate, respectively.  
In Eq.~(\ref{eq:color-pj}), the $t_A$, with $A=1,2, ..., N_c^2-1$ are the generators in the fundamental representation of the group SU($N_c$) color, and the subscripts, $ba$ and $dc$, represent 
the color indices of the heavy quark pair in the amplitude and those of its complex 
conjugate, respectively, but with $a,b,c,d = 1,2, ..., N_c$.

From Eq.~(\ref{eq:qqb2QQbgc}), calculating the short-distance hard part, 
$d\hat{\sigma}^{(3)}_{q+\bar{q}\to [Q\bar{Q}(\kappa)](p)}$,
is effectively the same as calculating the partionic cross section,
$d\sigma^{(3)}_{q+\bar{q}\to [Q\bar{Q}(\kappa)](p)}$, with the divergent gluon propagator 
of momentum $p$ replaced by its contact contribution.
From the normalization defined by the factorization formalism in Eqs.~(\ref{eq:pqcd_fac0}) and
(\ref{eq:pqcd_fac_h}), we obtain the expression for the NLP short-distance hard part as
\begin{eqnarray}
E_p \frac{d\hat{\sigma}^{(3)}_{q+\bar{q}\to [Q\bar{Q}(\kappa)](p)}}{d^3p}
=
\frac{1}{2\hat{s}}\left|\overline{\cal M}_{q\bar{q}\to[Q\bar{Q}(\kappa)]}\right|_c^2
\frac{1}{8\pi^2}\, \delta(\hat{s}+\hat{t}+\hat{u})\, ,
\label{eq:hardpart_def}
\end{eqnarray}
where $1/2\hat{s}$ is the partonic flux factor, 
$\left|\overline{\cal M}_{q\bar{q}\to[Q\bar{Q}(\kappa)]}\right|_c^2$ is the partonic scattering
amplitude squared with the initial-state spin and color averaged and 
the spin-color state of the final-state heavy quark pair defined by 
the projection operators in Eqs.~(\ref{eq:spin-pj}) and (\ref{eq:color-pj}), and 
where subscript ``$c$'' again indicates the use of the contact term of the divergent gluon propagator
of momentum $p$. Once more, this is equivalent to the removal of the gluonic pole contribution from
the gluon with a short bar in Fig.~\ref{fig:qq2QQg}, replacing the full propagator by the contact term, $G_c^{\rho\sigma}(p )$, Eq. (\ref{eq:gluonsp}).  
In Eq.~(\ref{eq:hardpart_def}), the last factor including the $\delta$-function is 
from the two-particle phase space, with the differential element $d^3p/E_p$ moved to the left of the equation.  
The parton-level Mandelstam variables in Eq.~(\ref{eq:hardpart_def}) are defined as
\begin{eqnarray}
\hat{s} = (p_1+p_2)^2\, , \quad
\hat{t} = (p_1-p)^2\, , \quad
\mbox{and}\ \quad
\hat{u} = (p_2-p)^2\, ,
\label{eq:stu-def}
\end{eqnarray}
with $\hat{s}+\hat{t}+\hat{u}=0$ imposed by the $\delta$-function. 
From Eq.~(\ref{eq:hardpart_def}), calculating the NLP partonic hard parts 
at ${\cal O}(\alpha_s^3)$ is  equivalent to calculating the spin-color
averaged partonic scattering matrix element square with the fragmenting gluonic 
pole contribution removed.

For convenience, we introduce a slightly simplified partonic hard part, ${H}$, 
to isolate the common factors for all scattering channels, 
\begin{eqnarray}
E_p \frac{d\hat{\sigma}^{(3)}_{q+\bar{q}\to [Q\bar{Q}(\kappa)](p)}}{d^3p}
\equiv 
\left[\frac{4\pi\alpha_s^3}{\hat{s}}\right]
\frac{1}{\bar{u} u \bar{v} v} \,
{H}_{q\bar{q}\to[Q\bar{Q}(\kappa)]}(\hat{s},\hat{t},\hat{u})\,
\delta(\hat{s}+\hat{t}+\hat{u})\, 
\label{eq:hardpart_eq} 
\end{eqnarray}
where the functions ${H}$ are defined by by
\begin{equation}
{H}_{q\bar{q}\to[Q\bar{Q}(\kappa)]}(\hat{s},\hat{t},\hat{u})
= \left|\overline{\cal M}_{q\bar{q}\to[Q\bar{Q}(\kappa)]}\right|_c^2
\left[ \frac{\bar{u} u \bar{v} v}{g_s^6}\right] \, .
\label{eq:hardpart_H}
\end{equation}
with coupling constant $g_s$.  The factors $u$, $\bar{u}$, $v$ and $\bar{v}$ 
are light-cone momentum fractions of heavy quark momenta 
$P_Q$ and $P_{\bar{Q}}$ of the scattering amplitude and 
$P'_Q$ and $P'_{\bar{Q}}$ in its complex conjugate, respectively, 
\cite{Kang:2014tta},
\begin{eqnarray}
P_Q = \frac{p}{2} + q_1 = u\, p = \frac{1+\zeta_1}{2}\, p\, , 
& \quad & 
P_{\bar{Q}} = \frac{p}{2} - q_1 = \bar{u}\, p = \frac{1-\zeta_1}{2}\, p\, ,
\nonumber\\
P'_Q = \frac{p}{2} + q_2 = v\, p = \frac{1+\zeta_2}{2}\, p\, , 
& \quad & 
P'_{\bar{Q}} = \frac{p}{2} - q_2 = \bar{v}\, p = \frac{1-\zeta_2}{2}\, p\, .
\label{eq:pqpqb}
\end{eqnarray}
Here, alternate variables $\zeta_1$ and $\zeta_2$ represent the light-cone momentum fraction flow 
between the heavy quark pair in the scattering amplitude and its complex conjugate, respectively.
Although the total momentum of the heavy quark pair in the amplitude and its complex
conjugate is the same, $P_Q+P_{\bar{Q}}=P'_Q+P'_{\bar{Q}}=p$, the relative momenta
between the pair, $q_1$ in the amplitude and $q_2$ in the complex conjugate amplitude, 
need not be the same.  That is, $\zeta_1$ (or $u$) does not have to equal  
$\zeta_2$ (or $v$), while $u+\bar{u} = 1$ and $v+\bar{v} = 1$.

The expression in Eq.~(\ref{eq:hardpart_H}) is actually valid for calculating 
the hard parts of all partonic scattering channels, including quark-gluon
and gluon-gluon scattering channels at ${\cal O}(\alpha_s^3)$.

\subsection{Short-distance coefficient for a heavy quark pair in an axial-vector spin state}
\label{subsec:hp_a}

For calculating the short-distance coefficients, or hard parts of the partonic process, 
$q(p_1)+\bar{q}(p_2)\to [Q\bar{Q}(\kappa)](p) + g$ with $n=a1$ and $a8$, 
we only need to consider  two diagrams, (a) and (b) in Fig.~\ref{fig:qq2QQg}.  
The other three diagrams in the figure vanish 
because of the $\gamma_5$ in the axial-vector spin projection operators
in Eq.~(\ref{eq:spin-pj}).  

\begin{figure}[!htp]
\noindent
\includegraphics[width=0.4\columnwidth]{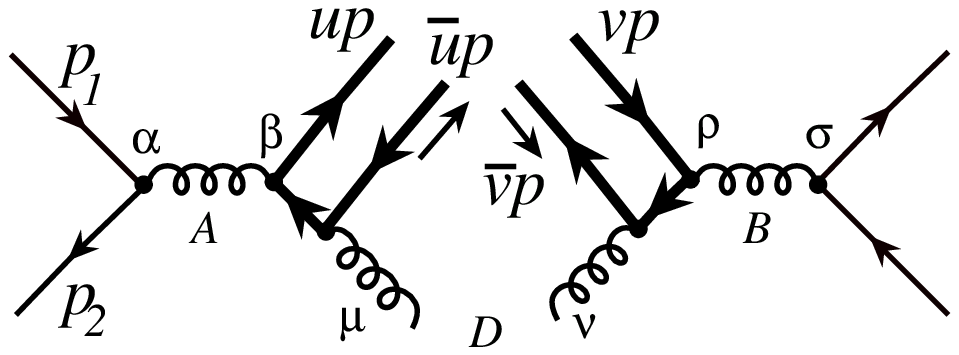}
\caption{Square of the diagram (a) in Fig.~\ref{fig:qq2QQg}.}
\label{fig:qqb2QQbg_sq}
\end{figure}

The only difference between producing a color singlet and a color octet heavy quark pair 
in an axial-vector spin state, $[Q\bar{Q}(a1)]$ vs $[Q\bar{Q}(a8)]$, is the color factor.  
For producing a color singlet pair, we find that the four terms from the square of two 
diagrams (a) and (b) in Fig.~\ref{fig:qq2QQg} have the same color factor, 
which can be derived from the square of diagram (a), 
as shown in Fig.~\ref{fig:qqb2QQbg_sq},
\begin{eqnarray}
C^{[1]} &=& \left(\frac{1}{N_c}\right)^2 \sum_{A,B,D}
{\rm Tr}\left[t^A t^B\right] 
\frac{1}{\sqrt{N_c}} {\rm Tr}\left[t^A t^D\right]
\frac{1}{\sqrt{N_c}} {\rm Tr}\left[t^B t^D\right]
\nonumber\\
&=&
\frac{N_c^2 - 1}{8 N_c^3}\, ,
\label{eq:color_a1}
\end{eqnarray}
where $(1/N_c)^2$ is from the average of initial-state quark and antiquark color, 
the $1/\sqrt{N_c}$ factor is from the definition of the color projection operator in Eq.~(\ref{eq:color-pj}), 
and all color indices are from the labels in Fig.~\ref{fig:qqb2QQbg_sq}.  
Unlike the color singlet case, the color factor for producing a color octet heavy quark pair 
in an axial-vector spin state is not the same for all four terms from the square of the two
diagrams.   From Fig.~\ref{fig:qqb2QQbg_sq}, we find for color factors $C^{[8]}_{ij^\dagger}$, 
with $i$ and $j$
labelling diagrams in the figure,
\begin{eqnarray}
C^{[8]}_{aa^\dagger} 
& = &  \left(\frac{1}{N_c}\right)^2 \sum_{A,B,D}
{\rm Tr}\left[t^A t^B\right] 
\sqrt{2}\, {\rm Tr}\left[t^C t^A t^D\right]
\sqrt{2}\, {\rm Tr}\left[t^C t^D t^B \right]
\nonumber\\
& = &
\left[ \frac{N_c^2 - 1}{8 N_c^3} \right] 
\left(N_c^2-2\right) \, ,
\label{eq:color_a8}
\end{eqnarray}
where the $\sqrt{2}$ factor is from the definition of the color projection operator in Eq.~(\ref{eq:color-pj}), 
and generator $t^C$ projects the octet state of the produced heavy quark pair,
and is summed over.  Similarly, we find the color factor for other three terms from
the squares of the diagrams (a) and (b) in Fig.~\ref{fig:qq2QQg},
\begin{eqnarray}
C^{[8]}_{ab^\dagger} 
& = &
- \left[ \frac{N_c^2 - 1}{4 N_c^3} \right] \, ,
\nonumber\\
C^{[8]}_{ba^\dagger} 
& = &
- \left[ \frac{N_c^2 - 1}{4 N_c^3} \right] 
= C^{[8]}_{ab^\dagger} \, ,
\nonumber\\
C^{[8]}_{bb^\dagger} 
& = &
\left[ \frac{N_c^2 - 1}{8 N_c^3} \right] 
\left(N_c^2-2\right) 
= C^{[8]}_{aa^\dagger} \, .
\label{eq:color_a8}
\end{eqnarray}
If we introduce two separate color factors,
\begin{equation}
{\cal C}_1 = \left[ \frac{N_c^2 - 1}{8 N_c} \right] ,
\quad \mbox{and} \quad
{\cal C}_2 = - \left[ \frac{N_c^2 - 1}{4 N_c^3} \right] ,
\label{eq:color_12}
\end{equation}
we have
\begin{eqnarray}
C^{[8]}_{aa^\dagger} = C^{[8]}_{bb^\dagger} = {\cal C}_1+{\cal C}_2,
\quad \mbox{and} \quad
C^{[8]}_{ab^\dagger} = C^{[8]}_{ba^\dagger} = {\cal C}_2\, .
\label{eq:color_a8_12}
\end{eqnarray}

For producing a color singlet axial-vector heavy quark pair, $[Q\bar{Q}(a1)]$, 
the amplitude squared of diagram (a) in Fig.~\ref{fig:qq2QQg}, 
as shown in Fig.~\ref{fig:qqb2QQbg_sq}, is given by
\begin{eqnarray}
\left|\overline{\cal M}^{aa^\dagger}_{q\bar{q}\to[Q\bar{Q}(a1)]}\right|^2
&=&
C^{[1]}\, g_s^6 \left(\frac{1}{2}\right)^2 
{\rm Tr}\left[\gamma\cdot p_1 \gamma^\sigma \gamma\cdot p_2 \gamma^\alpha\right]
\frac{(-g_{\alpha\beta})}{(p_1+p_2)^2} \frac{(-g_{\sigma\rho})}{(p_1+p_2)^2}
\nonumber\\
& \ & \times
{\rm Tr}\left[\gamma\cdot p\gamma_5 \gamma^\beta 
\frac{\gamma\cdot(u p -p_1-p_2)}{(up-p_1-p_2)^2}\gamma^\mu\right]
\nonumber\\
& \ & \times
{\rm Tr}\left[\gamma\cdot p\gamma_5 \gamma^\nu
\frac{\gamma\cdot(v p -p_1-p_2)}{(vp-p_1-p_2)^2}\gamma^\rho \right] (-g_{\mu\nu})
\nonumber\\
&=&
C^{[1]} \left[\frac{g_s^6}{\bar{u} \bar{v}}\right]\,
\frac{4}{\hat{s}}\left[\frac{\hat{t}^2 + \hat{u}^2 }{\hat{s}^2} \right]\, ,
\label{eq:qq2QQg_aa}
\end{eqnarray}
where the Feynman gauge was used for this gauge-invariant quantity.
Similarly, we find the other three terms from the squares of the diagrams (a) and (b)
in Fig.~\ref{fig:qq2QQg},
\begin{eqnarray}
\left|\overline{\cal M}^{ab^\dagger}_{q\bar{q}\to[Q\bar{Q}(a1)]}\right|^2
&=&
C^{[1]} \left[\frac{g_s^6}{\bar{u} {v}}\right]\,
\frac{4}{\hat{s}}\left[\frac{\hat{t}^2 + \hat{u}^2 }{\hat{s}^2} \right]\, ,
\label{eq:qq2QQg_ab}\\
\left|\overline{\cal M}^{ba^\dagger}_{q\bar{q}\to[Q\bar{Q}(a1)]}\right|^2
&=&
C^{[1]} \left[\frac{g_s^6}{{u} \bar{v}}\right]\,
\frac{4}{\hat{s}}\left[\frac{\hat{t}^2 + \hat{u}^2 }{\hat{s}^2} \right]\, ,
\label{eq:qq2QQg_ba}\\
\left|\overline{\cal M}^{bb^\dagger}_{q\bar{q}\to[Q\bar{Q}(a1)]}\right|^2
&=&
C^{[1]} \left[\frac{g_s^6}{{u} {v}}\right]\,
\frac{4}{\hat{s}}\left[\frac{\hat{t}^2 + \hat{u}^2 }{\hat{s}^2} \right]\, .
\label{eq:qq2QQg_bb}
\end{eqnarray}
Combining the four terms in Eqs.~(\ref{eq:qq2QQg_aa}), (\ref{eq:qq2QQg_ab}), 
(\ref{eq:qq2QQg_ba}), and (\ref{eq:qq2QQg_bb}), and using an identity satisfied by the momentum
fractions defined in Eq.\ (\ref{eq:pqpqb}), 
\begin{equation}
\frac{1}{\bar{u} \bar{v}} + \frac{1}{\bar{u} {v}} +
\frac{1}{{u} \bar{v}} + \frac{1}{{u} {v}}  
= \frac{1}{\bar{u} u \bar{v} v}  \, ,
\end{equation}
we derive the spin-color averaged matrix element squared for the partonic 
channel, $q(p_1)+\bar{q}(p_2)\to [Q\bar{Q}(a1)](p) + g$, as
\begin{equation}
\left|\overline{\cal M}_{q\bar{q}\to[Q\bar{Q}(a1)]}\right|^2
= 
\left[\frac{g_s^6}{\bar{u} u \bar{v} v}\right]\,
4C^{[1]} \left[\frac{\hat{t}^2 + \hat{u}^2 }{\hat{s}^3} \right]\, ,
\label{eq:qq2QQg_a1}
\end{equation}
where we have suppressed the subscript ``$c$'' for the squared matrix element, because no LP subtraction is necessary for these diagrams.
From the definition of the modified hard part in Eq.~(\ref{eq:hardpart_H}), we have
\begin{equation}
{H}_{q\bar{q}\to[Q\bar{Q}(a1)]}(\hat{s},\hat{t},\hat{u})
=4 \left[ \frac{N_c^2 - 1}{8 N_c^3} \right]
\left[\frac{\hat{t}^2 + \hat{u}^2 }{\hat{s}^3} \right]\, .
\label{eq:qq2QQa1_H}
\end{equation}


Since the spinor trace and the contraction of Lorentz indices are independent of the color, 
we derive the partonic scattering matrix element square for producing a $[Q\bar{Q}(a8)]$ pair as,
\begin{eqnarray}
\left|\overline{\cal M}^{aa^\dagger}_{q\bar{q}\to[Q\bar{Q}(a8)]}\right|^2
&=&
C_{aa^\dagger}^{[8]} \left[\frac{g_s^6}{\bar{u} \bar{v}}\right]\,
\frac{4}{\hat{s}}\left[\frac{\hat{t}^2 + \hat{u}^2 }{\hat{s}^2} \right]\, ,
\label{eq:qq2QQg_ab8}\\
\left|\overline{\cal M}^{ab^\dagger}_{q\bar{q}\to[Q\bar{Q}(a8)]}\right|^2
&=&
C_{ab^\dagger}^{[8]} \left[\frac{g_s^6}{\bar{u} {v}}\right]\,
\frac{4}{\hat{s}}\left[\frac{\hat{t}^2 + \hat{u}^2 }{\hat{s}^2} \right]\, ,
\label{eq:qq2QQg_ab8}\\
\left|\overline{\cal M}^{ba^\dagger}_{q\bar{q}\to[Q\bar{Q}(a8)]}\right|^2
&=&
C_{ba^\dagger}^{[8]} \left[\frac{g_s^6}{{u} \bar{v}}\right]\,
\frac{4}{\hat{s}}\left[\frac{\hat{t}^2 + \hat{u}^2 }{\hat{s}^2} \right]\, ,
\label{eq:qq2QQg_ba8}\\
\left|\overline{\cal M}^{bb^\dagger}_{q\bar{q}\to[Q\bar{Q}(a8)]}\right|^2
&=&
C_{bb^\dagger}^{[8]} \left[\frac{g_s^6}{{u} {v}}\right]\,
\frac{4}{\hat{s}}\left[\frac{\hat{t}^2 + \hat{u}^2 }{\hat{s}^2} \right]\, .
\label{eq:qq2QQg_bb8}
\end{eqnarray}
Combining all four terms together, recognizing 
\begin{eqnarray}
   \frac{1}{\bar{u} \bar{v}}C_{aa^\dagger}^{[8]}  
+ \frac{1}{\bar{u} {v}}C_{ab^\dagger}^{[8]} 
+ \frac{1}{{u} \bar{v}}C_{ba^\dagger}^{[8]} 
+ \frac{1}{{u} {v}}C_{bb^\dagger}^{[8]} 
=\frac{1}{\bar{u} u \bar{v} v} \left[
\frac{1}{2}(1+\zeta_1 \zeta_2)\, {\cal C}_1 + {\cal C}_2 \right] ,
\label{eq:color_a8}
\end{eqnarray}
and using ${\cal C}_1$ and ${\cal C}_2$ from Eq.~(\ref{eq:color_12}), we obtain
\begin{equation}
{H}_{q\bar{q}\to[Q\bar{Q}(a8)]}(\hat{s},\hat{t},\hat{u})
=2 \left[ \frac{N_c^2 - 1}{8 N_c} \right]
\left[1 + \zeta_1 \zeta_2 - \frac{4}{N_c^2} \right]
\left[\frac{\hat{t}^2 + \hat{u}^2 }{\hat{s}^3} \right]\, ,
\label{eq:qq2QQa8_H}
\end{equation}
where $\zeta_1$ and $\zeta_2$ are heavy quark momentum fractions defined in 
Eq.~(\ref{eq:pqpqb}).

\subsection{Short-distance coefficient for a heavy quark pair production with vector spin}
\label{subsec:hp_v8}

For producing a color singlet heavy quark pair from quark-antiquark scattering
at ${\cal O}(\alpha_s^3)$, only diagrams (a) and (b) in Fig.~\ref{fig:qq2QQg} contribute
since the other three diagrams can only produce the pair in a color octet state.
Since the color is independent of the spin state of the pair,
the color factor for producing a color singlet pair in a vector spin state
is the same as $C^{[1]}$ in Eq.~(\ref{eq:color_a1}).

Similar to Eq.~(\ref{eq:qq2QQg_aa}), we have from the diagram in Fig.~\ref{fig:qqb2QQbg_sq}, 
\begin{eqnarray}
\left|\overline{\cal M}^{aa^\dagger}_{q\bar{q}\to[Q\bar{Q}(v1)]}\right|^2
&=&
C^{[1]}\, g_s^6 \left(\frac{1}{2}\right)^2 
{\rm Tr}\left[\gamma\cdot p_1 \gamma^\sigma \gamma\cdot p_2 \gamma^\alpha\right]
\frac{(-g_{\alpha\beta})}{(p_1+p_2)^2} \frac{(-g_{\sigma\rho})}{(p_1+p_2)^2}
\nonumber\\
& \ & \times
{\rm Tr}\left[\gamma\cdot p \gamma^\beta 
\frac{\gamma\cdot(u p -p_1-p_2)}{(up-p_1-p_2)^2}\gamma^\mu\right]
\nonumber\\
& \ & \times
{\rm Tr}\left[\gamma\cdot p \gamma^\nu
\frac{\gamma\cdot(v p -p_1-p_2)}{(vp-p_1-p_2)^2}\gamma^\rho \right] (-g_{\mu\nu})
\nonumber\\
&=&
C^{[1]} \left[\frac{g_s^6}{\bar{u} \bar{v}}\right]\,
\frac{4}{\hat{s}}\left[\frac{\hat{t}^2 + \hat{u}^2 }{\hat{s}^2} \right]\, ,
\label{eq:qq2QQg_vaa}
\end{eqnarray}
which is the same as $\left|\overline{\cal M}^{aa^\dagger}_{q\bar{q}\to[Q\bar{Q}(a1)]}\right|^2$.
In the same way, we find for the other three terms contributing to the production of a $[Q\bar{Q}(v1)]$ pair,
\begin{eqnarray}
\left|\overline{\cal M}^{ab^\dagger}_{q\bar{q}\to[Q\bar{Q}(v1)]}\right|^2
&=&
- C^{[1]} \left[\frac{g_s^6}{\bar{u} {v}}\right]\,
\frac{4}{\hat{s}}\left[\frac{\hat{t}^2 + \hat{u}^2 }{\hat{s}^2} \right]\, ,
\label{eq:qq2QQg_vab}\\
\left|\overline{\cal M}^{ba^\dagger}_{q\bar{q}\to[Q\bar{Q}(v1)]}\right|^2
&=&
- C^{[1]} \left[\frac{g_s^6}{{u} \bar{v}}\right]\,
\frac{4}{\hat{s}}\left[\frac{\hat{t}^2 + \hat{u}^2 }{\hat{s}^2} \right]\, ,
\label{eq:qq2QQg_vba}\\
\left|\overline{\cal M}^{bb^\dagger}_{q\bar{q}\to[Q\bar{Q}(v1)]}\right|^2
&=&
C^{[1]} \left[\frac{g_s^6}{{u} {v}}\right]\,
\frac{4}{\hat{s}}\left[\frac{\hat{t}^2 + \hat{u}^2 }{\hat{s}^2} \right]\, ,
\label{eq:qq2QQg_vbb}
\end{eqnarray}
where the interference terms have the opposite sign compared
to the corresponding terms for producing a $[Q\bar{Q}(a1)]$ pair, 
Eqs.\ (\ref{eq:qq2QQg_ab}) and (\ref{eq:qq2QQg_ba}).  
Combining Eqs.~(\ref{eq:qq2QQg_vaa}), (\ref{eq:qq2QQg_vab}), 
(\ref{eq:qq2QQg_vba}), and (\ref{eq:qq2QQg_vbb}), we obtain
\begin{equation}
{H}_{q\bar{q}\to[Q\bar{Q}(v1)]}(\hat{s},\hat{t},\hat{u})
=4 \left[ \frac{N_c^2 - 1}{8 N_c^3} \right]
\left[\frac{\hat{t}^2 + \hat{u}^2 }{\hat{s}^3} \right]\, \zeta_1\zeta_2 \, ,
\label{eq:qq2QQv1_H}
\end{equation}
which differs from ${H}_{q\bar{q}\to[Q\bar{Q}(a1)]}(\hat{s},\hat{t},\hat{u})$ only by 
an overall factor of $\zeta_1\zeta_2$, and vanishes if the produced heavy quark and 
antiquark have the same momentum.


For the production of a color octet heavy quark pair in a vector spin state 
from quark and antiquark scattering at ${\cal O}(\alpha_s^3)$, all five diagrams in 
Fig.~\ref{fig:qq2QQg} can contribute.  
With three more diagrams for producing a $[Q\bar{Q}(v8)]$ pair, 
each combination of the diagrams in the scattering amplitude and its complex conjugate 
has its unique color factor, labeled  $C_{ij^\dagger}^{[8]}$ for diagram ``$i$'' 
multiplied by the complex conjugate diagram ``$j$'' with $i,j=$(a), (b), (c), (d), and (e) 
 in Fig.~\ref{fig:qq2QQg}.  We find, however, that all of these twenty-five color 
factors can be expressed in terms of the two color factors, 
${\cal C}_1$ and ${\cal C}_2$, as defined in Eq.~(\ref{eq:color_12}), and we 
present all of them in Table~\ref{tab:color}.   To normalize the color factor involving 
the three-gluon vertex, we take the following convention for the Feynman rule of 
the three-gluon vertex.  For the three-gluon vertex of diagram (c) in Fig.~\ref{fig:qq2QQg}, 
we let $-g_s f^{EAD} = (-ig_s) (-i f^{EAD})$, and include the ``$(-i f^{EAD})$'' 
into the calculation of the color factor, while keeping ``$(-ig)$'' with the calculation of 
the rest of diagram.  We follow the same convention for the three-gluon vertices 
in the complex conjugate of the scattering amplitude.
\begin{table}[h]
\caption{\label{tab:color} 
Color factors for all combinations of diagrams in the scattering amplitude and 
its complex conjugate for the partonic process, $q+\bar{q} \to [Q\bar{Q}(v8)] + g$, 
expressed in terms of ${\cal C}_1$ and ${\cal C}_2$ defined in Eq.~(\ref{eq:color_12}).}
\begin{tabular}{c|c|c|c|c|c}
\hline\hline \itshape 
~$C^{[8]}_{ij^\dagger}$~ 
      & (a) & (b) & (c) & (d) & (e) \\ \hline 
(a) & ~${\cal C}_1 +  {\cal C}_2$~  &  ${\cal C}_2$ &  $- {\cal C}_1$ 
     & ${\cal C}_1 +  {\cal C}_2$ &  ${\cal C}_2$ \\ \hline
(b) & $ {\cal C}_2$  &  ~${\cal C}_1+ {\cal C}_2$~  &  ${\cal C}_1$ 
     &  ${\cal C}_2$ & ${\cal C}_1 +  {\cal C}_2$ \\ \hline
(c) & ~$ - {\cal C}_1$~  &  ${\cal C}_1 $  &  $2\, {\cal C}_1$ 
     &  $- {\cal C}_1$ & ${\cal C}_1$ \\ \hline
(d) & $ {\cal C}_1 + {\cal C}_2 $  &  ${\cal C}_2 $  &  $-{\cal C}_1$ 
     &  ~${\cal C}_1 + {\cal C}_2/2 $~ & ${\cal C}_2/2$ \\ \hline
(e) & $ {\cal C}_2 $  &  ${\cal C}_1 + {\cal C}_2 $  &  ${\cal C}_1$ 
     &  ${\cal C}_2/2 $ & ~${\cal C}_1 +{\cal C}_2/2$~ \\ 
\hline\hline
\end{tabular}
\end{table}

From our discussion leading to Eq.~(\ref{eq:hardpart_H}), 
we need to calculate the initial-state spin-color averaged scattering amplitude square,
$\left|\overline{\cal M}_{q\bar{q}\to[Q\bar{Q}(v8)]}\right|_c^2$, 
 for a quark and an antiquark to produce a heavy $[Q\bar{Q}(v8)]$ pair with 
 the LP gluonic pole contribution removed.
That is, we need to use the contact term of the gluon propagator 
for the gluon with a short bar and momentum $p$ in Fig.~\ref{fig:qq2QQg}.  
Although three particles are produced in the final state, the scattering process,
$q(p_1)+\bar{q}(p_2)\to[Q\bar{Q}(v8)](p) + g(k)$, has effectively a ``$2\to 2$'' 
kinematics, and has three independent external momenta due to 
momentum conservation, $p_1 + p_2 = p + k$.
With the use of the contact term for the gluon propagator, Eq.~(\ref{eq:gluonsp}),
$i \hat{n}^\mu \hat{n}^\nu/(p\cdot \hat{n})^2$, 
the calculated partionic hard parts for this production channel, 
$q(p_1)+\bar{q}(p_2)\to[Q\bar{Q}(v8)](p) + g(k)$, can depend on the ratios, 
$p_1\cdot \hat{n}/p\cdot \hat{n}$ and $p_2\cdot \hat{n}/p\cdot \hat{n}$,
if we choose $p_1$, $p_2$, and $p$ as the three independent momentum vectors 
for this partonic subprocess. 

In Eq.~(\ref{eq:gluonsp}), the auxiliary vector $\hat{n}^\mu$ is defined to 
be conjugate to the heavy quark pair momentum $p^\mu$, in the sense defined after Eq. (\ref{eq:cv_g0}).  
Since the squares of the diagrams 
in Fig.~\ref{fig:qq2QQg}, $\left|\overline{\cal M}_{q\bar{q}\to[Q\bar{Q}(v8)]}\right|^2$, 
are Lorentz invariant, and the subtraction term that removes the LP contribution in
Eq.~(\ref{eq:qqb2QQb}) is Lorentz invariant, the NLP hard part defined in
Eq.~(\ref{eq:hardpart_H}) is Lorentz invariant as well.  
That is, with the ``$2\to 2$'' kinematics, the NLP hard parts 
can be expressed in terms of the Mandelstam variables 
defined in Eq.~(\ref{eq:stu-def}), as can the ratios, 
$p_1\cdot \hat{n}/p\cdot \hat{n}$ and $p_2\cdot \hat{n}/p\cdot \hat{n}$.
Since the inner products, $p_1\cdot \hat{n}$, $p_2\cdot \hat{n}$, and $p\cdot \hat{n}$, 
are Lorentz scalars, we can evaluate them in any Lorentz frame.  
The most convenient choice of $\hat n^\mu$ makes the gluon polarization 
in Eq.~(\ref{eq:cv_g0}) transverse in
the center of mass frame of the parton-parton scattering,
in which the heavy quark pair moves along the $z$-axis, and 
the unobserved final-state parton of momentum 
$k^\mu=p_1^\mu+p_2^\mu-p^\mu$ with $k^2=0$ 
moves along the $-z$-axis, back-to-back to the $p^\mu$.  In this frame, 
the unobserved final-state parton momentum $k^\mu$ should be exactly 
proportional to the auxiliary vector $\hat{n}^\mu$.  
Since the contact term of gluon propagator is invariant when the vector $\hat{n}^\mu$ is rescaled, 
we can write the contact term of gluon propagator of momentum $p$ in this frame as
\begin{equation}
\frac{i\, \hat{n}^\mu\, \hat{n}^\nu}{(p\cdot \hat{n})^2}  = \frac{i\, k^\mu\, k^\nu}{(p\cdot k)^2}\, .
\label{eq:gluonsp_k}
\end{equation}
Thus, we can replace $\hat{n}^\mu$ by $k^\mu$ for our calculation of the NLP hard parts
in this frame.  The Lorentz invariant ratios involving the $\hat{n}^\mu$ vector 
can be then expressed in terms of the Mandelstam variables as
\begin{equation}
\frac{p_1\cdot \hat{n}}{p\cdot \hat{n}} = \frac{-\hat{u}}{\hat{s}} \,
\quad\quad
\frac{p_2\cdot \hat{n}}{p\cdot \hat{n}} = \frac{-\hat{t}}{\hat{s}} \, .
\label{eq:p1p2pn}
\end{equation}
The Lorentz invariance of these ratios and of the scattering amplitude  
ensures that the NLP hard parts evaluated in this frame are valid in all other 
Lorentz frames, including the center of mass frame of the colliding hadrons
or the lab frame.
This simplification is specific to the $2\to 2$ subprocess that 
characterizes this calculation of the NLP cross section at LO.

With the Feynman diagrams in Fig.~\ref{fig:qq2QQg}, spin projection operators in
Eq.~(\ref{eq:spin-pj}), the contact gluon propagator in Eq.~(\ref{eq:gluonsp_k}), 
and the calculated color factors in Table~\ref{tab:color}, it is straightforward to 
derive the hard parts for partonic production channel, 
$q(p_1)+\bar{q}(p_2)\to[Q\bar{Q}(v8)](p) + g(k)$.  
Since all color factors from combinations of the diagrams in Fig.~\ref{fig:qq2QQg} 
and its complex conjugate can be expressed in terms of two color factors,
${\cal C}_1$ and ${\cal C}_2$, as shown in Table~\ref{tab:color}, 
we express our calculated short-distance hard part of this channel as
\begin{equation}
{H}_{q\bar{q}\to[Q\bar{Q}(v8)]}(\hat{s},\hat{t},\hat{u})
=
2 \left[ \frac{N_c^2-1}{8N_c} \right]\, {\cal H}_1(\hat{s},\hat{t},\hat{u})
+ 
4 \left[ -\frac{N_c^2-1}{4N_c^3} \right]\, {\cal H}_2(\hat{s},\hat{t},\hat{u})\, ,
\label{eq:qq2QQv8_H}
\end{equation}
where we find that
\begin{eqnarray}
{\cal H}_1 
&=&
\frac{\hat{t}^2 + \hat{u}^2}{\hat{s}^3}\left[
(1+\zeta_1\zeta_2)\zeta_1\zeta_2+4(1-\zeta_1^2)(1-\zeta_2^2)\right]
+ 
\frac{\hat{t}^2 - \hat{u}^2}{\hat{s}^3}\left[
(1-\zeta_1\zeta_2)(\zeta_1+\zeta_2)\right]
\nonumber\\
&\ & 
-
\frac{(\hat{t} - \hat{u})^2}{\hat{s}^3}\left[
\zeta_1^2(1-\zeta_2^2)+\zeta_2^2(1-\zeta_1^2)\right]\, ,
\nonumber \\
{\cal H}_2 
&=&
\frac{\hat{t}^2 + \hat{u}^2}{\hat{s}^3}
\left[\zeta_1\zeta_2\right]
+
\frac{\hat{t}^2 - \hat{u}^2}{\hat{s}^3}\left[
\zeta_1(1-\zeta_2^2)+\zeta_2(1-\zeta_1^2)\right]
+
\frac{1}{\hat{s}}\left[
(1-\zeta_1^2)(1-\zeta_2^2)\right]\, .
\label{eq:qq2QQv8_H1H2}
\end{eqnarray}
This result for the $(v8)$ heavy pair color-spin configuration, 
together with Eqs.\ (\ref{eq:qq2QQa1_H}), (\ref{eq:qq2QQa8_H}) and 
(\ref{eq:qq2QQv1_H}) for the $(a1)$, $(a8)$ and $(v1)$ configurations, respectively,
gives the full ${\cal O}(\alpha_s^3)$ contribution to heavy pair production in the light pair channel.
The partonic short-distance hard part for producing a heavy quark pair in a tensor spin state 
from the scattering of a light quark pair vanishes at the order of $\alpha_s^3$.  This is simply
because  the tensor spin projection operator in Eq.~(\ref{eq:spin-pj}) 
has an even number of Dirac $\gamma$ matrices, and the trace of odd number of $\gamma$-matrices 
vanishes.

Using essentially the same methods described in this section, we have calculated the NLP 
short-distance hard parts for other partonic scattering channels at ${\cal O}(\alpha_s^3)$,
including $q+g \to [Q\bar{Q}(\kappa)]+q$, $g+q \to [Q\bar{Q}(\kappa)]+q$, 
and $g+g\to [Q\bar{Q}(\kappa)]+g$, with the produced heavy quark pair in 
all possible spin-color states. The complete results of perturbatively calculated hard parts 
for all partonic scattering channels are presented in Appendix~\ref{app:hardparts}.

\newpage
\section{Predictive power and connection to NRQCD}
\label{sec:nrqcd-ffs}

Even with the first non-trivial order of partonic hard parts calculated in this paper, and
additional future improvement of the hard parts with higher order perturbative corrections, 
the predictive power of the QCD collinear factorization formalism in Eq.~(\ref{eq:pqcd_fac0}) 
for heavy quarkonium production still relies on knowledge of the nonperturbative, 
but universal, heavy quarkonium fragmentation functions: $D_{f\to H}(z,\mu_F^2;m_Q)$ and 
${\cal D}_{[Q\bar{Q}(\kappa)]\to H}(z,u,v,\mu_F^2;m_Q)$.
As derived in Ref.~\cite{Kang:2014tta}, these universal heavy quarkonium FFs
satisfy a closed set of evolution equations that determines their dependence on the factorization scale, $\mu_F$.   
The first non-trivial order of all evolution kernels is also available in Ref.\ \cite{Kang:2014tta}.  
Higher order corrections to these evolution kernels could be systematically calculated 
in perturbation theory.  That is, we are able, in principle, to derive all heavy quarkonium 
FFs at any factorization scale $\mu_F$ with a set of input distributions at 
an initial factorization scale, $\mu_0 \gtrsim 2m_Q$, 
at which the power suppressed contribution in $1/\mu_F^2$ is compatible with
the leading logarithmic contribution in $\ln\mu_F^2$.   
Like all other QCD factorization formalisms, the predictive power 
of the factorization formula for heavy quarkonium production in Eq.~(\ref{eq:pqcd_fac0}) 
thus depends on a set of nonperturbative input FFs 
(as well as the input PDFs for hadronic collisions).  
Since both the short-distance partonic hard parts and the evolution kernels of 
these FFs are perturbatively calculated, it is the FFs at input scale $\mu_0$ that
are the most sensitive to the detailed properties of the heavy quarkonia produced, 
including their spin and angular momenta.
If we were to follow the procedure familiar for light parton PDFs and FFs, 
we would need to extract heavy quarkonium non-perturbative input FFs
from experimental data or possible lattice QCD calculations,
similar to the extraction of light hadron, such as pion or proton, FFs.

However, unlike the light hadron FFs, heavy quarkonium FFs at the input scale
$\mu_0 \gtrsim 2m_Q$ depend on a large perturbative scale - the heavy quark mass,
$m_Q\sim {\cal O}(\mu_0) \gg \Lambda_{\rm QCD}$.  Even more importantly, 
this perturbative scale is substantially separated from the momentum scale 
needed for the binding of heavy quarkonium, $m_Q v$ and any other non-perturbative 
scales of the bound state.  With such a clear separation of momentum scales, 
NRQCD is a natural effective theory of QCD to separate
the dynamics of the fragmentation process at the perturbative scale $\mu_0\sim {\cal O}(m_Q)$ 
from the physics at the scale $m_Q v$ and below.
Appealing to NRQCD factorization for this production process as a very plausible conjecture,
we can, at least as a reasonable model,  
express all heavy quarkonium FFs in terms of the universal NRQCD 
long-distance matrix elements (LDMEs)
with perturbatively calculated functional dependence on the momentum 
fractions, $z$, $\zeta_1$, and $\zeta_2$.  
In this approach, all input heavy quarkonium FFs are expanded in terms of the LDMEs, 
according to their effective powers in heavy quark's velocity $v$ 
in the heavy quark pair's rest frame. 
These LDMEs are the same as those used in light parton FFs to heavy quarkonia, so that
this approach has the attractive feature of introdcing no new nonperturbative parameters 
relative to those that are already in the LP expansion \cite{Bodwin:2014gia}.
 The perturbatively 
calculated coefficient for each LDME is further expanded in terms of the power of 
strong coupling constant $\alpha_s(\mu_0)$.  
With the small value of the velocity, $v$, the perturbative expansion
of all heavy quarkonium FFs can be expressed in terms of a very small number of universal
LDMEs for each physical quarkonium state to enhance the predictive power
of the QCD factorization formalism in Eq.~(\ref{eq:pqcd_fac0}) tremendously.

\subsection{NRQCD factorization and input fragmentation functions}
\label{subsec:input-ffs}

The NRQCD factorization approach to heavy quarkonium production was 
proposed to express the inclusive cross section for the direct production 
of a quarkonium state $H$  as a sum of ``short-distance'' coefficients times 
NRQCD LDMEs \cite{Bodwin:1994jh}, 
\begin{equation}
\sigma^H(p_T,m_Q) = \sum_{[Q\bar{Q}(n)]} 
\hat{\sigma}_{[Q\bar{Q}(n)]}(p_T,m_Q,\mu_\Lambda)
\langle 0| {\cal O}_{[Q\bar{Q}(n)]}^H(\mu_\Lambda)|0\rangle \, ,
\label{eq:nrqcd-fac}
\end{equation}
where $p_T$ is the transverse momentum of produced heavy quarkonium, and 
$\mu_\Lambda\sim {\cal O}(m_Q)$ is the ultraviolet cut-off of the 
NRQCD effective theory, or effectively, is the factorization scale of the factorization 
formalism.  When the physical scale $p_T\gg m_Q$, it was demonstrated 
in Ref.~\cite{Kang:2014tta} that the perturbative expansion of the short-distance 
coefficient functions, $\hat{\sigma}_{[Q\bar{Q}(n)]}(p_T,m_Q,\mu_\Lambda)$
in Eq.~(\ref{eq:nrqcd-fac}), in powers of $\alpha_s$ is not always stable
since high orders in $\alpha_s$ can be enhanced by the powers
of $p_T/m_Q$, as well as powers of large logarithms in $\ln(p_T/m_Q)$.  
The QCD factorization formalism in Eq.~(\ref{eq:pqcd_fac0}) was in fact 
proposed to reorganize both of these large power and logarithmic enhancements.

With the input factorization scale, $\mu_0 \sim {\cal O}(m_Q)$, and the large
momentum scale separation between $\mu_0$ and 
all other nonperturbative scales of the heavy quarkonium, we propose as above and 
as a conjecture, to use NRQCD factorization for calculating the heavy quarkonium
FFs at the input scale $\mu_0$ as,
\begin{equation}
D_{f\to H}(z,\mu_0^2;m_Q)
= \sum_{[Q\bar{Q}(n)]} 
\hat{d}_{f\to [Q\bar{Q}(n)]}(z,\mu_0^2;m_Q,\mu_\Lambda) 
\langle {\cal O}_{[Q\bar{Q}(n)]}^H(\mu_\Lambda)\rangle \, ,
\label{eq:frag_fac1}
\end{equation}
for the heavy quarkonium FFs from a single parton of flavor $f=q,\bar{q},g$, and
\begin{equation}
{\cal D}_{[Q\bar{Q}(\kappa)]\to H}(z,u,v,\mu_0^2;m_Q)
= \sum_{[Q\bar{Q}(n)]} 
\hat{d}_{[Q\bar{Q}(\kappa)]\to [Q\bar{Q}(n)]}(z,u,v,\mu_0^2;m_Q,\mu_\Lambda)
\langle {\cal O}_{[Q\bar{Q}(n)]}^H(\mu_\Lambda)\rangle 
\label{eq:frag_fac2q}
\end{equation}
for the heavy quarkonium FFs from a perturbatively produced heavy quark 
pair of the spin-color state $\kappa$ defined in Sec.~\ref{sec:hardparts}.
In Eq.~(\ref{eq:frag_fac1}), $\hat{d}_{f\to [Q\bar{Q}(n)]}(z,\mu_0^2;m_Q,\mu_\Lambda)$ 
is the perturbatively calculable short-distance coefficient function for an
off-shell parton of flavor $f$ to evolve into a non-relativistic heavy quark
pair represented by $[Q\bar{Q}(n)]$ with $n$ expressed in terms of the 
standard spectroscopic notation, $^{2S+1}L_J^{[1,8]}$, according to the 
spin $S$, orbital angular momentum $L$, and total angular momentum $J$, 
as well as the color state ($[1]$ for singlet and $[8]$ for octet) of the pair. 
Similarly, $\hat{d}_{[Q\bar{Q}(\kappa)]\to [Q\bar{Q}(n)]}(z,u,v,\mu_0^2;m_Q,\mu_\Lambda)$ 
is the short-distance coefficient functions for an off-shell perturbatively 
produced heavy quark pair with the spin-color quantum number $\kappa$ to 
evolve into the same non-relativistic heavy quark pair state $[Q\bar{Q}(n)]$.

Since the short-distance coefficient functions, 
$\hat{d}_{f\to [Q\bar{Q}(n)]}(z,\mu_0^2;m_Q,\mu_\Lambda)$ in Eq.~(\ref{eq:frag_fac1}),
and $\hat{d}_{[Q\bar{Q}(\kappa)]\to [Q\bar{Q}(n)]}(z,u,v,\mu_0^2;m_Q,\mu_\Lambda)$
in Eq.~(\ref{eq:frag_fac2q}), are not sensitive to the details of the produced heavy 
quarkonium $H$, they can be  calculated systematically by projecting the factorization
formalisms in Eqs.~(\ref{eq:frag_fac1}) and (\ref{eq:frag_fac2q}) on 
a pair of non-relativistic heavy quarks of mass $m_Q$. 
Note that the heavy quarkonium fragmentation functions defined in both
Eqs.~(\ref{eq:frag_fac1}) and (\ref{eq:frag_fac2q}) are boost invariant along 
the direction of heavy quarkonium momentum of $p^\mu$.
Let $[Q\bar{Q}(c)]$ be such a state with $c$ expressed in terms of the standard 
spectroscopic notation.  The corresponding projection operators for such a non-relativistic 
heavy quark pair are given in the Appendix~A of Ref.~\cite{Ma:2013yla}, 
or in the references therein.  
By expanding both sides of the factorization formulas in Eqs.~(\ref{eq:frag_fac1}) 
and (\ref{eq:frag_fac2q}) order-by-order in powers of $\alpha_s$, 
the short-distance coefficient functions can be perturbatively extracted 
by calculating both sides in perturbation theory.  

For example, by expanding the factorization formula in Eq.~(\ref{eq:frag_fac2q})
for producing a non-relativistic heavy quark pair, $[Q\bar{Q}(\COaSz)]$,
to zeroth order in power of $\alpha_s$, we have \cite{Ma:2013yla},
\begin{equation}
{\cal D}^{(0)}_{[Q\bar{Q}(\kappa)]\to [Q\bar{Q}(\COaSz)]}(z,u,v)
= 
\hat{d}^{(0)}_{[Q\bar{Q}(\kappa)]\to [Q\bar{Q}(\COaSz)]}(z,u,v)
\label{eq:dfac-lo}
\end{equation}
where the superscript ``$(0)$''  indicates 
the zeroth order in power of $\alpha_s$, 
the dependence on the heavy quark mass and factorization scales are suppressed,
and the normalization of NRQCD LDMEs, 
$\langle {\cal O}_{[Q\bar{Q}(n)]}^{[Q\bar{Q}(\COaSz)]}(0)\rangle^{(0)}=1$
with $n=\COaSz$, is used.
The perturbatively produced heavy quark pair states, $[Q\bar{Q}(\kappa)]$, 
are defined in terms of relativistic heavy quark field operators in QCD 
with the vector, axial-vector, and tensor spin states and singlet and octet color states,  
along with the projection operators defined in Ref.~\cite{Kang:2014tta}.  
On the other hand, the NRQCD states of a heavy quark pair are defined in terms of 
non-relativistic heavy quark fields of NRQCD.  
As a result, there can be non-trivial matching coefficients even at zeroth order
in  $\alpha_s$.  With our definition of heavy quarkonium FFs and the normalization
of NRQCD LDMEs, we have, for example,
\begin{equation}
\label{eq:resLOS}
\hat{d}^{\text{(0)}}_{[Q\bar{Q}(a8)]\to [Q\bar{Q}(\COaSz)]}(z,u,v)
=\frac{1}{N_c^2-1}\frac{1}{2m_Q}\,\delta(1-z)\,
\delta(2u-1)\,\delta(2v-1),
\end{equation}
where $u=(1-\zeta_1)/2$ and $v=(1-\zeta_2)/2$.  A complete list of the zeroth order
short-distance coefficient functions can be found in Ref.~\cite{Ma:2013yla}.

Beyond the LO in $\alpha_s$, the partonic FFs to a 
non-relativistic heavy quark pair on the left-hand side of the factorization 
formulas in both Eqs.~(\ref{eq:frag_fac1}) and (\ref{eq:frag_fac2q}) 
have several types of perturbative divergences, as do the NRQCD LDMEs
to a non-relativistic heavy quark pair on the right-hand side of these equations.  
With the finite heavy quark mass, $m_Q$, 
the partonic FFs on the left-hand side of Eqs.~(\ref{eq:frag_fac1}) and (\ref{eq:frag_fac2q}) 
have no collinear  (CO) divergence.  
The ultraviolet (UV) divergence associated with the composite operators 
defining the FFs are systematically removed by the UV counter-terms (UVCT),
required as a necessary part of the definition of these FFs, while the UV divergences 
associated with the virtual loop diagrams are taken care of by the standard renormalization
of QCD perturbation theory.  The factorization scheme, associated with the 
cancelation of the UV divergence between the partonic FFs and the UVCT should
be chosen to be the same as the factorization scheme used to calculate 
the  short-distance partonic hard parts of the QCD collinear factorization formalism 
in Eq.~(\ref{eq:pqcd_fac0}).  For calculating the partonic FFs for producing 
a heavy quark pair in a non-relativistic $S$-wave state, the infrared (IR) divergence 
associated with contributions from individual Feynman diagrams completely cancels
at any given order of $\alpha_s$ after we sum up all contributions at this order.  
However, for calculating the partonic FFs of producing a heavy quark pair in a non-relativistic 
$P$-wave or higher orbital angular momentum state,  IR divergences (as well as what is often referred 
to as the rapidity divergence 
\cite{Collins:2008ht,Chiu:2009yx,Becher:2011dz,Chiu:2012ir,Echevarria:2013aca} 
in the context of transverse momentum dependent 
factorization formalism \cite{Collins:1981uk}) cannot be completely 
canceled by summing over contributions from all diagrams \cite{Ma:2014eja}.  
Instead,  IR divergences (and the rapidity divergences) should be cancelled by
corresponding divergences in the NRQCD LDMEs on the RHS 
of the factorization formalism, as required by  factorization.
In addition to the UV and IR divergences, the partonic FFs for 
producing a pair of heavy quarks has Coulomb divergences
from the exchange of soft gluons between the pair.  
The Coulomb divergence of the partonic FFs on the LHS of 
Eqs.~(\ref{eq:frag_fac1}) and (\ref{eq:frag_fac2q}) should be exactly 
cancelled by the Coulomb divergence of the NRQCD LDMEs on the RHS 
to ensure the validity of the factorization.  
Although there is no formal proof for the NRQCD factorization formalisms 
in Eqs.~(\ref{eq:frag_fac1}) and (\ref{eq:frag_fac2q}), the NLO calculation
of the short-distance hard parts for both the single parton and heavy quark pair
FFs in Refs.~\cite{Ma:2013yla,Ma:2014eja} confirms that 
all UV, IR (as well as rapidity), and Coulomb divergences
are completely cancelled to leave all hard parts at this order infrared safe (IRS).
NRQCD factorization effectively predicts the functional dependence of the
heavy quarkonium FFs in terms of momentum fractions: $z$, $u$, and $v$ 
(or equivalently $z$, $\zeta_1$ and $\zeta_2$), and NRQCD LDMEs up to the
approximation to truncate the perturbative expansion in powers of $\alpha_s$
and $v$ in Eqs.~(\ref{eq:frag_fac1}) and (\ref{eq:frag_fac2q}) \cite{Ma:2014svb}. 

It is the input FFs that are the most sensitive to the characteristics of individual 
heavy quarkonium produced in high energy scattering, since both the short-distance
partonic hard parts in the QCD collinear factorization formalism in Eq.~(\ref{eq:pqcd_fac0})
and the evolution kernels of the FFs are perturbatively calculated and completely
universal for the production of any heavy quarkonium states.  
The input FFs determine the difference in the production various
states, including their spin and polarization dependencee, as well as the normalization of 
their production rates.  The QCD factorization in Eq.~(\ref{eq:pqcd_fac0}) assures
that these input FFs are universal regardless of whether the heavy quarkonium state 
is produced in a hadron-hadron, lepton-hadron or lepton-lepton collisions.  
In summary, input FFs are essential for understanding the characteristic 
differences between all heavy quarkonium states produced.
  
\subsection{Relation between QCD factorization and NRQCD factorization}
\label{subsec:matching}

The NRQCD factorization for the input FFs in Eqs.~(\ref{eq:frag_fac1}) and 
(\ref{eq:frag_fac2q}) is independent of the QCD collinear factorization in Eq.~(\ref{eq:pqcd_fac0}).
The two factorizations have their own power counting and perturbative expansions.  
The QCD factorization is valid up to the first power corrections in the $1/p_T^2$ expansion 
of the cross section, while the short-distance hard parts and evolution kernels of FFs 
can be systematically improved by calculating higher order corrections in powers 
of $\alpha_s$.   As noted above, although the NRQCD factorization has not been fully 
proved perturbatively for high orders in powers $\alpha_s$ and $v$, explicit NLO 
calculation up to $v^4$ in LDMEs verifies the factorization.  
In order to best compare with experimental data, it is important to get the most accurate 
calculations from each factorization series, and in particular, the NRQCD factorization
for the input FFs to test the universality of these functions.

If the NRQCD factorization for the input FFs in Eqs.~(\ref{eq:frag_fac1}) and 
(\ref{eq:frag_fac2q}) is valid to all orders in  $\alpha_s$ and the relative 
heavy quark velocity $v$, the validity of the QCD collinear factorization formalism 
for heavy quarkonium production in Eq.~(\ref{eq:pqcd_fac0}) effectively ensures 
that the NRQCD factorization for the production cross section in Eq.~(\ref{eq:nrqcd-fac})
is valid at least for the  LP and NLP terms in the $1/p_T$ expansion.  In addition, the QCD factorization formalism 
reorganizes the perturbative expansion of NRQCD factorization 
by resumming all powers of  $\ln(p_T/m_Q)$.
However, this equivalence between QCD factorization and NRQCD factorization
does not say anything about the validity of the NRQCD factorized cross section 
in Eq.~(\ref{eq:nrqcd-fac}) beyond the NLP terms.

In terms of QCD collinear factorization in Eq.~(\ref{eq:pqcd_fac0}), all partonic
hard parts for production cross sections and evolution kernels of the FFs are calculated 
with the heavy quark mass neglected.  The heavy quark mass dependence of 
the NRQCD factorization for the production cross section can be systematically 
included by the following perturbative matching formalism,
\begin{eqnarray}
E_P\frac{d\sigma_{A+B\to H+X}}{d^3P}(P,m_Q)
&\equiv & E_P\frac{d\sigma_{A+B\to H+X}^{\rm QCD}}{d^3P}(P,m_Q=0)
\label{eq:matching}\\
&+ &
E_P\frac{d\sigma_{A+B\to H+X}^{\rm NRQCD}}{d^3P}(P,m_Q\neq 0)
- E_P\frac{d\sigma_{A+B\to H+X}^{\rm QCD-Asym}}{d^3P}(P,m_Q=0)\, ,
\nonumber
\end{eqnarray}
where $\sigma^{\rm QCD}$ is given in Eq.~(\ref{eq:pqcd_fac0}) with the input FFs
calculated in NRQCD factorization, $\sigma^{\rm NRQCD}$ is given by Eq.~(\ref{eq:nrqcd-fac})
with only the LP and NLP terms in the $1/p_T$ expansion, and the ``asymptotic" $\sigma^{\rm QCD-Asym}$ is 
defined to be the same as  $\sigma^{\rm QCD}$ with the FFs expanded to a fixed order
in both $\alpha_s$ and $v$ to match the order used to calculate $\sigma^{\rm NRQCD}$.  
If the NRQCD factorization formalism in Eq.~(\ref{eq:nrqcd-fac}) is valid beyond the NLP, 
the $\sigma^{\rm NRQCD}$ in Eq.~(\ref{eq:matching}) should include terms beyond the NLP.
In Eq.~(\ref{eq:matching}), the first term on the RHS is more reliable for  large $p_T$, while 
the second term is more suited for the low $p_T$ region, and the third term effectively removes
the double counting at any given fixed order in power of $\alpha_s$.  
Consequently, the combined formula in Eq.~(\ref{eq:matching}) could be consistent with 
experimental measurement of heavy quarkonium cross sections for a wider range of 
$p_T (> m_Q)$. 

\subsection{An example}
\label{subsec:singlet}

In this subsection, we provide an explicit example to demonstrate how the very large 
and complex NLO contribution to the heavy quarkonium production calculated 
in the color singlet model (also in NRQCD) can be reproduced by a much simpler and 
fully analytic LO calculation in terms of QCD factorization, Eq.~(\ref{eq:pqcd_fac0}) 
with the FFs calculated in NRQCD.  The same comparison for other partonic production channels 
can be found in Ref.~\cite{Ma:2014svb}.  

From the QCD factorization formalism in Eq.~(\ref{eq:pqcd_fac0}) and the NRQCD factorization 
formalisms in Eqs.~(\ref{eq:frag_fac1}) and (\ref{eq:frag_fac2q}), we found that the LO contribution 
to the production of a color singlet spin-1 heavy quark pair, $[Q\bar{Q}(\CScSa)]$, 
which matches to a physical heavy quarkonium $H$ by a NRQCD LDME, 
$\langle {\cal O}_{[Q\bar{Q}(\CScSa)]}^H\rangle$, is given by the combination of 
producing a color-octet heavy quark pair, which fragments into a color singlet and 
spin-1 NRQCD state, $[Q\bar{Q}(\CScSa)]$.  At order of $\alpha_s$, only the pair in a 
vector $[Q\bar{Q}(v8)]$ or an axial-vector $[Q\bar{Q}(a8)]$ spin state can fragment into
the spin-1 NRQCD state~\footnote{The contribution from the vector spin state, 
$[Q\bar{Q}(v8)]$, was not included in our previous short paper Ref.~\cite{Kang:2011mg}.}.
The LO partonic hard parts at ${\cal O}(\alpha_s^3)$ for producing a heavy quark pair 
in both $[Q\bar{Q}(v8)]$ and $[Q\bar{Q}(a8)]$ perturbative states are given 
in the Appendix~\ref{app:hardparts} found from the detailed of derivation given 
in Sec.~\ref{sec:hardparts}.  The ${\cal O}(\alpha_s)$ heavy quarkonium FFs, 
${\cal D}^{(1)}_{[Q\bar{Q}(\kappa)]\to [Q\bar{Q}(\CScSa)]\to H}$ with $\kappa=a8,v8$, 
can be calculated by using Eq.~(\ref{eq:frag_fac2q}).  

\bigskip
\begin{figure}[!htp]
\noindent
\begin{minipage}[c]{0.4\columnwidth}
\noindent
        \includegraphics[width=0.7\columnwidth]{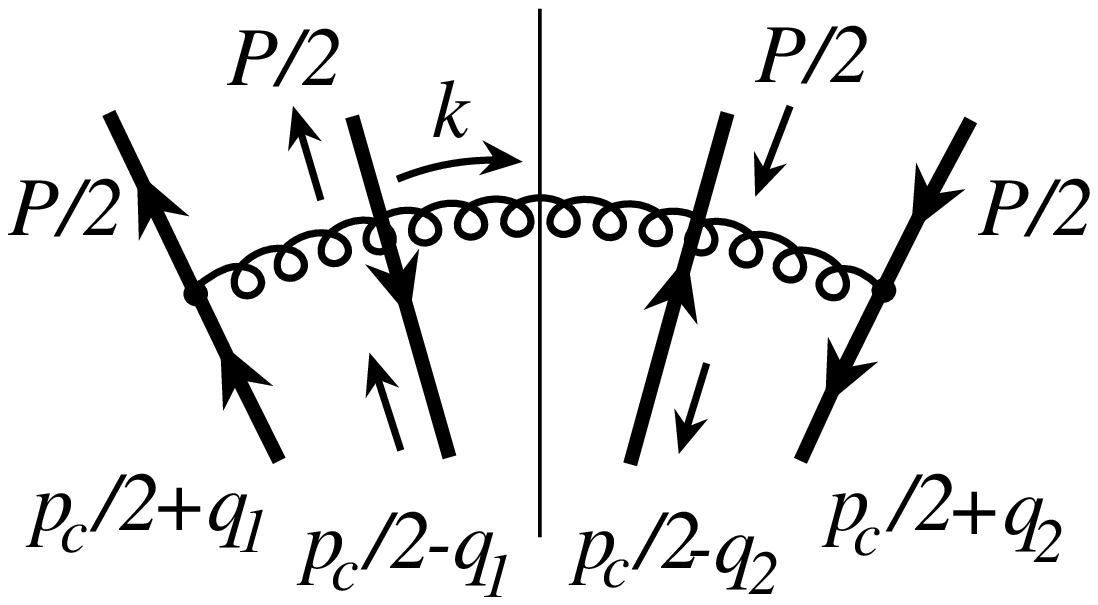}
\end{minipage}
\begin{minipage}[c]{0.03\columnwidth}
+
\end{minipage}
\begin{minipage}[c]{0.4\columnwidth}
\noindent
        \includegraphics[width=0.7\columnwidth]{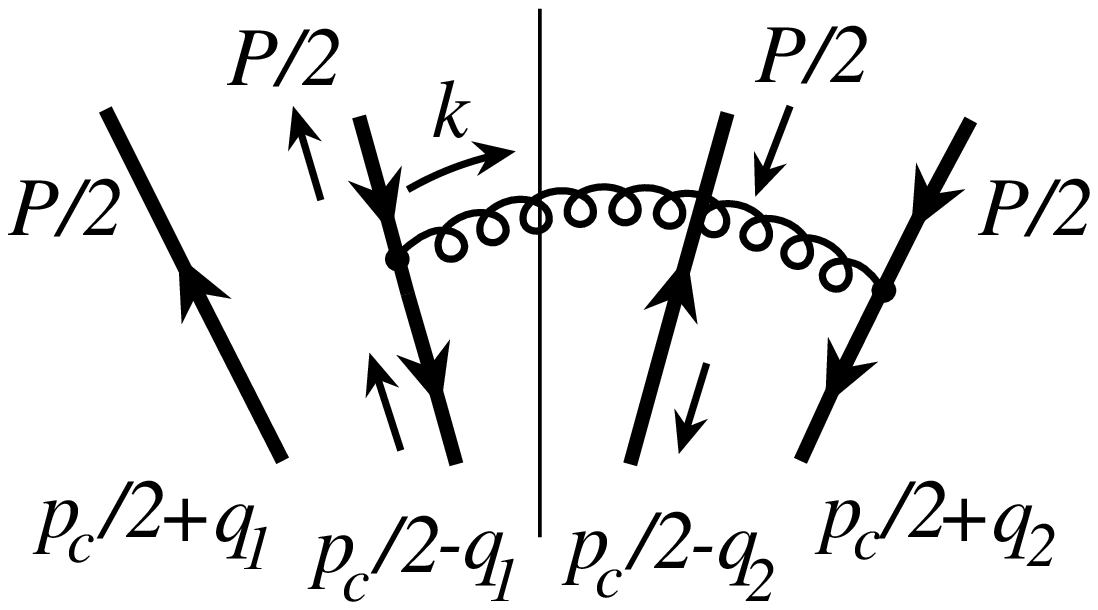}
\end{minipage}
\begin{minipage}[c]{0.08\columnwidth}
+ ...
\end{minipage}
\caption{Leading order Feynman diagrams representing the fragmentation of a heavy quark pair to another heavy quark pair.}
\label{fig:fragfunc}
\end{figure}

At order of $\alpha_s$, the relevant Feynman diagrams for calculating the heavy quarkonium
FFs, ${\cal D}^{(1)}_{[Q\bar{Q}(\kappa)]\to [Q\bar{Q}(\CScSa)]\to H}$ with $\kappa=a8,v8$,
in NRQCD are given in Fig.~\ref{fig:fragfunc}, where  the amplitude and its complex conjugate 
are combined together in the cut diagram notation.  As in the standard NRQCD calculation,
the momenta of the heavy quark and antiquark (upper lines of the diagrams) are fixed 
at $p/2$ and Dirac indices of these lines are contracted with an NRQCD 
singlet spin-1 projection operator \cite{Ma:2013yla}.  
The only difference for the fragmentation between a vector $[Q\bar{Q}(v8)]$
and an axial-vector $[Q\bar{Q}(a8)]$ state is that the lower fragmenting heavy quark and antiquark
lines in Fig.~\ref{fig:fragfunc} are contracted with different projection operators, as defined  
in Eq.~(\ref{eq:spin-pj}).  We obtain,
\begin{eqnarray}
&&
{\cal D}^{\text{CR}}_{[Q\bar{Q}(v8)]\to [Q\bar{Q}(\CScSa)]\to H}(z, u, v, \mu^2; m_Q)
=
\frac{1}{2N_c^2}\frac{\langle {\cal O}^{H}_{[Q\bar{Q}(\CScSa)]}\rangle}{3m_Q}\,
\nonumber\\
&& {\hskip 0.5 in}
\times
\frac{\alpha_s}{2\pi} \,
\Delta_{-}(u, v)\,
\frac{z}{1-z}
\left[
\ln\left(r(z) + 1\right)
+\left(1-4z+2z^2\right)
\left(1- \frac{1}{1+r(z)}\right)
\right],
\label{eq:frag-v8}
\\
&&
{\cal D}^{\text{CR}}_{[Q\bar{Q}(a8)]\to [Q\bar{Q}(\CScSa)]\to H}(z, u, v, \mu^2; m_Q)
=
\frac{1}{2N_c^2}\frac{\langle {\cal O}^{H}_{[Q\bar{Q}(\CScSa)]}\rangle}{3m_Q}\,
\nonumber\\
&& {\hskip 0.5 in}
\times
\frac{\alpha_s}{2\pi}\,
\Delta_{+}(u, v)\,
z(1-z)
\left[
\ln\left(r(z) + 1\right)
+ \left(1- \frac{1}{1+r(z)}\right)
\right],
\label{eq:frag-a8}
\end{eqnarray}
for the fragmentation of a $[Q\bar{Q}(v8)]$ state and of a $[Q\bar{Q}(a8)]$ state, respectively.  
In Eqs.~(\ref{eq:frag-v8}) and (\ref{eq:frag-a8}), 
the superscript ``CR'' indicates the use of a cutoff regularization scheme
for the logarithmic UV divergence, 
the function $r(z) \equiv z^2\mu^2/(4m_Q^2(1-z)^2)$, and $\Delta_{\pm}(u, v)$ are defined as
\begin{eqnarray}
\Delta_+(u, v) &= & \frac{1}{4} \left[\delta\left(u-\frac{z}{2}\right) + \delta\left(\bar u-\frac{z}{2}\right)\right]
\left[ \delta\left(v-\frac{z}{2}\right) + \delta\left(\bar v-\frac{z}{2}\right)\right],
\nonumber\\
\Delta_-(u, v)&=&\frac{1}{4} \left[\delta\left(u-\frac{z}{2}\right) - \delta\left(\bar u-\frac{z}{2}\right)\right]
\left[ \delta\left(v-\frac{z}{2}\right) - \delta\left(\bar v-\frac{z}{2}\right)\right].
\label{eq:Delta_pm}
\end{eqnarray}
As explained in Sec.~(\ref{subsec:input-ffs}), UV counter-terms are needed to calculate the FFs perturbatively in order to renormalize the composite operators that define these heavy quarkonium FFs.   
In deriving Eqs.~(\ref{eq:frag-v8}) and (\ref{eq:frag-a8}), we renormalized the UV divergence by a cutoff $\mu^2$ on the transverse momentum integration, $dk_T^2$, which is directly connected to the virtuality of the fragmenting heavy quark pair.  The heavy quark mass, $m_Q$, effectively 
removes the potential CO divergence.  

The perturbatively calculated FFs in Eqs.~(\ref{eq:frag-v8}) and (\ref{eq:frag-a8}) are not unique
due to the renormalization of the perturbative UV divergence.
Just like the light-hadron FFs, the exact expression of heavy quarkonium FFs depend on 
the factorization scheme, which is a direct consequence of the renormalization ambiguity 
for removing the perturbative UV divergence.
For a comparison, we also list here the same FFs calculated with dimensional regularization and
the $\overline{\rm MS}$ renormalization scheme~\cite{Ma:2013yla,Ma:2014eja}, 
\begin{eqnarray}
&&
{\cal D}^{\text{DR}}_{[Q\bar{Q}(v8)]\to [Q\bar{Q}(\CScSa)]\to H}(z, u, v, \mu^2; m_Q)
= \frac{1}{2N_c^2}\frac{\langle {\cal O}^{H}_{[Q\bar{Q}(\CScSa)]}\rangle}{3m_Q}
\nonumber\\
&& {\hskip 0.5in}
\times
\frac{\alpha_s}{2\pi} \,
\Delta_{-}(u, v)\,
\frac{z}{1-z}
\left[
\ln\left(\frac{\mu^2}{4(1-z)^2m_Q^2}\right)
+\left(1-4z+2z^2\right) 
\right],
\label{eq:fragDR-v8}
\\
&&
{\cal D}^{\text{DR}}_{[Q\bar{Q}(a8)]\to [Q\bar{Q}(\CScSa)]\to H}(z, u, v, \mu^2; m_Q)
= \frac{1}{2N_c^2}\frac{\langle {\cal O}^{H}_{[Q\bar{Q}(\CScSa)]}\rangle}{3m_Q}
\nonumber\\
&& {\hskip 0.5 in}
\times
\frac{\alpha_s}{2\pi}\,
\Delta_{+}(u, v)\,
z(1-z)
\left[
\ln\left(\frac{\mu^2}{4(1-z)^2m_Q^2}\right)
- 1 \right],
\label{eq:fragDR-a8}
\end{eqnarray}
for the fragmentation of a $[Q\bar{Q}(v8)]$ state and a $[Q\bar{Q}(a8)]$ state,
respectively.  In Eqs.~(\ref{eq:fragDR-v8}) and (\ref{eq:fragDR-a8}) 
the superscript ``DR'' indicates the use of  dimensional regularization.
The difference between the FFs in Eqs.~(\ref{eq:frag-v8}) and (\ref{eq:frag-a8})
and those in Eqs.~(\ref{eq:fragDR-v8}) and (\ref{eq:fragDR-a8}) reflects the 
factorization scheme dependence of the perturbatively calculated FFs, which should lead
to differences in fitted values of the NRQCD LDMEs, for example, the value of 
$\langle {\cal O}^{H}_{[Q\bar{Q}(\CScSa)]}\rangle$ in these equations.

From Eq.~(\ref{eq:Delta_pm}), it is clear that $\Delta_+(u, v)$ and $\Delta_-(u, v)$ have  
different symmetry properties under the transformation of 
$u\leftrightarrow \bar{u}=1-u$ or $v\leftrightarrow \bar{v}=1-v$,
or equivalently, $\zeta_1\leftrightarrow -\zeta_1$ or $\zeta_2\leftrightarrow -\zeta_2$:
$\Delta_+(u, v)$ is symmetric, while $\Delta_-(u, v)$ is anti-symmetric. 
That is, when the relative momentum fraction of the heavy quark pair in the amplitude
or in its complex conjugate, reverses its direction, the calculated FF, 
${\cal D}^{(1)}_{[Q\bar{Q}(a8)]\to [Q\bar{Q}(\CScSa)]\to H}(z, u, v, \mu^2; m_Q)$, which is
proportional to $\Delta_+(u, v)$ is symmetric, while 
${\cal D}^{(1)}_{[Q\bar{Q}(v8)]\to [Q\bar{Q}(\CScSa)]\to H}(z, u, v, \mu^2; m_Q)$ is anti-symmetric.
This symmetry property effectively requires that under the same transformation, 
only the symmetric part of partonic hard parts for producing a $[Q\bar{Q}(a8)]$ pair, 
or the antisymmetric part of partonic hard parts for producing a $[Q\bar{Q}(v8)]$ pair, 
can give a nonvanishing contribution to the production cross section at this order.  For example,
the LO QCD factorization contribution from Eq.~(\ref{eq:pqcd_fac0}) to 
hadronic $J/\psi$ production via a color singlet $[Q\bar{Q}(\CScSa)]$ channel
can be symbolically given by,
\begin{eqnarray}
d\sigma_{A+B\to J/\psi} 
&=& \sum_{ab} \phi_{A\to a}\otimes \phi_{B\to b} \left[ 
  d\hat{\sigma}^{A}_{ab\to [Q\bar{Q}(v8)]} \otimes 
  {\cal D}^{(1)}_{[Q\bar{Q}(v8)]\to [Q\bar{Q}(\CScSa)]\to J/\psi}
  \right.
  \nonumber\\
&\ & \hskip 1.2in \left.
+d\hat{\sigma}^{S}_{ab\to [Q\bar{Q}(a8)]} \otimes 
{\cal D}^{(1)}_{[Q\bar{Q}(a8)]\to [Q\bar{Q}(\CScSa)]\to J/\psi}
\right]\, ,
\label{eq:singlet-lo}
\end{eqnarray}
where $\sum_{a,b}$ sums over all possible initial-state parton flavors, 
and the superscripts, ``$A$'' and ``$S$'', represent the ``antisymmetric'' and ``symmetric'' 
property of the partonic hard parts under the transformation 
$\zeta_1\leftrightarrow -\zeta_1$ or $\zeta_2\leftrightarrow -\zeta_2$.

The relation between the partonic cross sections $d\hat{\sigma}$ and 
the short-distance hard parts, $H$, is given in Eq.~(\ref{eq:hardpart_eq}).
The complete results for short-distance hard parts at ${\cal O}(\alpha_s^3)$ 
are listed in the Appendix~\ref{app:hardparts}, and their derivation was given 
in Sec.~\ref{sec:hardparts}.  The symmetric part for producing 
a color-octet axial-vector pair, $[Q\bar{Q}(a8)]$, 
was derived in one of our previous papers on the subject~\cite{Kang:2011mg}, 
and is summarized here
\begin{eqnarray}
&&H_{q\bar q\to [Q\bar{Q}(a8)]g}^S
=\frac{(N_c^2-4)(N_c^2-1)}{4N_c^3}
\left[ \frac{(\hat t^2+\hat u^2)}{\hat s^3}\right] \, ,
\label{eq:qq2QQa8-s}\\
&&H_{gq \to [Q\bar{Q}(a8)]q}^S
= \frac{(N_c^2-4)}{4N_c^2}
\left[ \frac{(\hat s^2+\hat u^2)}{-\hat t^3}\right]\, ,
\\
&&H_{gg\to [Q\bar{Q}(a8)]g}^S
= \frac{(N_c^2-4)}{N_c^2-1}
\left[\frac{(-\hat{s}\hat{t}-\hat{t}\hat{u}-\hat{u}\hat{s})^3}{(\hat{s}\hat{t}\hat{u})^3}\right]\, ,
\end{eqnarray}
where the superscript ``$S$'' indicates keeping only the symmetric terms of the hard parts 
under the transformation of $\zeta_1\leftrightarrow -\zeta_1$ or $\zeta_2\leftrightarrow -\zeta_2$.

For the production of a color-octet vector pair, $[Q\bar{Q}(v8)]$, we need only the hard parts 
that are antisymmetric under the transformation of $\zeta_1\leftrightarrow -\zeta_1$ or $\zeta_2\leftrightarrow -\zeta_2$.  That is, only the terms that are of odd powers in both $\zeta_1$ and $\zeta_2$ are relevant.  From Eqs.~(\ref{eq:qq2QQv8_H}) and (\ref{eq:qq2QQv8_H1H2}), 
we obtain, 
\begin{eqnarray}
H_{q\bar q\to [Q\bar{Q}(v8)]g}^A
&=&\frac{(N_c^2-4)(N_c^2-1)}{4N_c^3}
\left[\frac{(\hat t^2+\hat u^2)}{\hat s^3}\right]\, \zeta_1\, \zeta_2
\label{eq:qq2QQv8-a}\\
&=&
\zeta_1\, \zeta_2\, H_{q\bar q\to [Q\bar{Q}(a8)]g}^S \, ,
\label{eq:a8vsv8}
\end{eqnarray}
where the superscript ``$A$'' indicates keeping the antisymmetric terms when 
$\zeta_1\leftrightarrow -\zeta_1$ or $\zeta_2\leftrightarrow -\zeta_2$.
From Appendix~\ref{app:hardparts}, we find that the relation in Eq.~(\ref{eq:a8vsv8}) 
is actually true for all partonic scattering channels at ${\cal O}(\alpha_s^3)$, 
\begin{eqnarray}
H_{ab\to [Q\bar{Q}(v8)]c}^A = (u-\bar u)(v-\bar v)H_{ab\to [Q\bar{Q}(a8)]c}^S\, ,
\end{eqnarray}
where the identities, $u-\bar{u}=\zeta_1$ and $v-\bar{v}=\zeta_2$ are used, 
and $a$, $b$ and $c$, run over all possible parton flavors: quark, antiquark and gluon.

\begin{figure}[!htp]
\includegraphics[width=0.7\columnwidth]{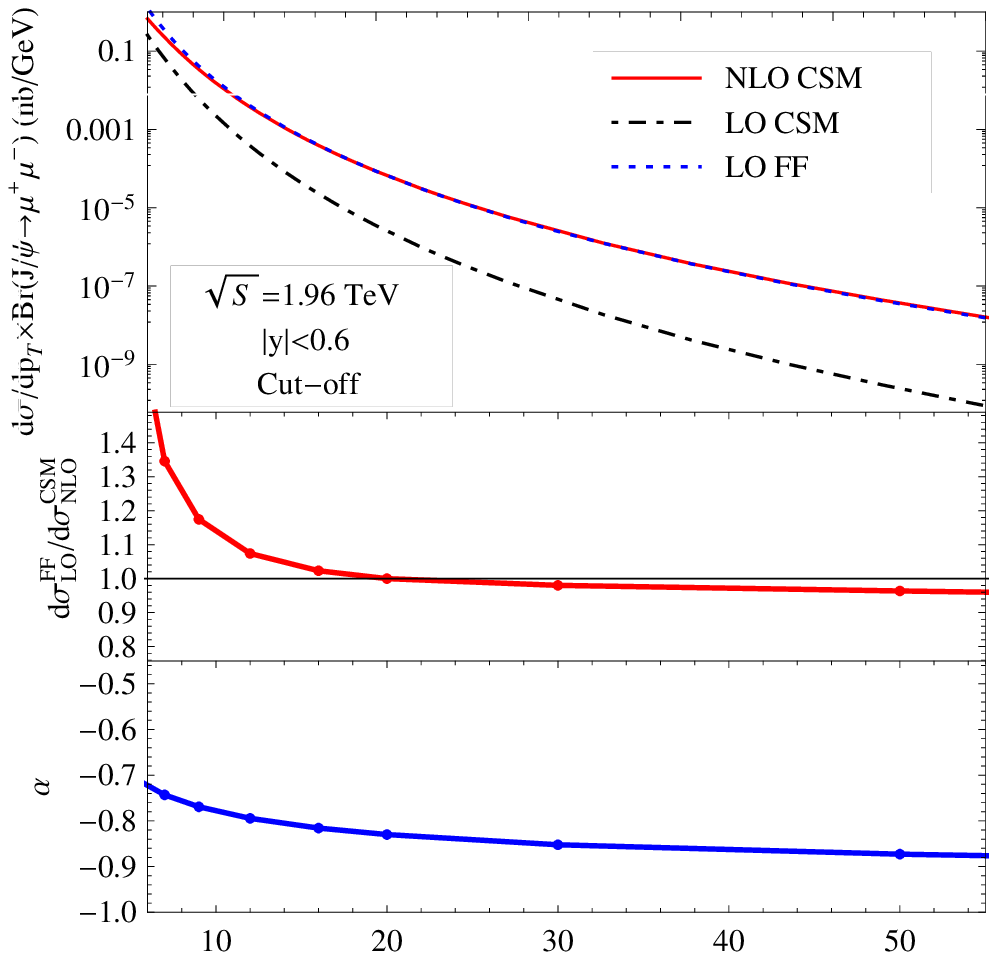}
\caption{Upper panel: comparison of LO QCD factorization results with FFs 
calculated in the ``CR'' - the cutoff regularization and renormalization scheme (solid line)
with the LO (dot-dashed line) and NLO (dashed line) results calculated in CSM. 
Middle panel: ratio of LO QCD factorization results over NLO CSM results. 
Lower panel: polarization parameter evaluated with the LO QCD factorization formalism
and FFs given in Eqs.~(\ref{eq:fragpol-v8}) and (\ref{eq:fragpol-a8}).}
\label{fig:cutoff}
\end{figure}

In Fig.~\ref{fig:cutoff}, we compare the NLO results of $J/\psi$ production 
(solid line), calculated in terms of the color singlet model (a special case of NRQCD), 
with the LO results of QCD factorization (dashed line).  Both results are for 
$J/\psi$ production multiplied by the branching ratio to a $\mu^+\mu^-$ pair 
and evaluated at the Tevatron energy at $\sqrt{S}=1.96$~TeV and 
in the central rapidity region with $|y|<0.6$.  
The NLO results of the CSM calculation are from Ref.~\cite{Campbell:2007ws}.
The LO QCD factorization results were generated by using 
Eq.~(\ref{eq:pqcd_fac0}), or more precisely, Eq.~(\ref{eq:singlet-lo}), 
for proton-antiproton collisions with the LO partonic hard parts and 
the perturbative FFs calculated in this section {\it without} the evolution (or resummation).
We set the factorization scale $\mu_F$  equal to the renormalization scale $\mu$, 
and $\mu_F=\mu=p_T$.   To be consistent with our factorized LO calculations, 
we used CTEQ6L1 parton distribution functions (PDFs) \cite{Pumplin:2002vw}, 
and the one-loop expression for $\alpha_s$ with $n_f=5$ active flavors and 
$\Lambda_{\text{QCD}}^{(5)}=165$~MeV.  In addition, 
we set the charm quark mass $m_Q=1.5$ GeV, and NRQCD
LDME, $\langle {\cal O}^{J/\psi}_{[Q\bar{Q}(\CScSa)]}\rangle=1.32~{\text{GeV}^3}$. 
From the upper and central panels of Fig.~\ref{fig:cutoff}, we find that 
our LO contribution to the production cross section, which is effectively a NLP
contribution from the QCD factorization formalism in Eq.~(\ref{eq:pqcd_fac0}), 
gives a nice description of the very complicated NLO CSM results 
when the $J/\psi$'s $p_T$ is sufficiently large.  In addition, 
as shown in Ref.~\cite{Ma:2014svb}, the QCD factorization formalism in
Eq.~(\ref{eq:pqcd_fac0}) with fully analytical LO short-distance hard parts 
and fully analytical NLO FFs calculated in NRQCD,
without higher order resummation or evolution, can reproduce the state of the art, 
numerical NLO NRQCD results, 
channel by channel for $p_T \gtrsim 10$~GeV.

In Fig.~\ref{fig:cutoff}, we have also plotted the LO results of the CSM 
(dot-dashed line), which is much more than an order of magnitude 
smaller than the NLO results of the CSM calculation.
This is exactly caused by the order $p_T^2/m_Q^2$ power enhancement that 
the NLO  has over the LO  in the CSM calculation 
(the same issue appears in NRQCD calculations as well) \cite{Kang:2014tta}.  
That is, the expansion of CSM in powers of $\alpha_s$ 
is not perturbatively reliable because some of the higher order terms are enhanced 
by large powers and/or logarithms of such large powers \cite{Kang:2014tta}. 
Although the LO QCD factorization results (solid line) 
are from the NLP terms ($\sim {\cal O}(1/p_T^6)$) of the QCD factorization 
formalism, they are still power enhanced in comparison to the LO CSM results, 
which are actually of the ${\cal O}(1/p_T^8)$ \cite{Kang:2014tta}.  
The key difference between the QCD factorization formalism in Eq.~(\ref{eq:pqcd_fac0})
and the NRQCD factorization formalism in Eq.~(\ref{eq:nrqcd-fac}) is that all 
perturbatively calculated short-distance hard parts and evolution kernels of 
the QCD factorization formalism are evaluated at a {\it single} hard scale 
so that they are free of any large higher order enhancement 
from the power of large momentum ratios or the logarithms of such large ratios.

\newpage
\section{Heavy quarkonium polarization}
\label{sec:pol}

Understanding the polarization of produced heavy quarkonia is critically important for 
determining the true QCD dynamics, as well as the mechanism of production.  
In terms of the QCD factorization formalism in Eq.~(\ref{eq:pqcd_fac0}), 
as pointed out earlier in this paper, all hadronic properties of the  heavy quarkonia,
including their spin and polarizations, are only sensitive to the FFs at the 
input scale $\mu_0\gtrsim 2m_Q$, since all perturbatively calculated partonic hard parts 
and evolution kernels of FFs are insensitive to the details of the produced  
states.  In this section, we adapt the same NRQCD factorization conjecture to evaluate the 
input FFs to a polarized heavy quarkonium state.  We introduce the basic method for 
calculating these input FFs, and present a complete example for the calculation of 
the input FFs to a polarized $J/\psi$ via a polarized color singlet and spin-1 
nonrelativistic heavy quark pair ($\CScSa$).  A complete set of polarized heavy quarkonium
FFs, and their derivation in detail, calculated to NLO in the NRQCD factorization approach 
including all possible partonic scattering channels at this order, can be found 
in Refs.~\cite{Ma:2013yla,Ma:2014eja,Zhang:thesis}.

If we keep only the LP term in QCD collinear factorization, Eq.~(\ref{eq:pqcd_fac0}),
for the production of a heavy quarkonium at $p_T\gg m_Q$, 
the hard partonic collision produces a single, energetic virtual parton state 
at the short-distance scale $\sim {\cal O}(1/p_T)$, followed by a fragmentation process 
to generate a physical $J/\psi$, represented by the FFs of the produced single parton.  
Although the single parton FFs to a heavy quarkonium are non-perturbative, 
their factorization scale dependence is given by the DGLAP evolution equations with
perturbatively calculated evolution kernels (expanded in powers of $\alpha_s$).  
The DGLAP  equations evolve the fragmentation process, 
initiated by the  energetic parton, {\it perturbatively}, to the input scale, 
$\mu_0\gtrsim 2m_Q$ or a distance scale $\sim {\cal O}(1/2m_Q)$, 
where a heavy quark pair emerges from the fragmenting parton,
and the pair eventually fragments into a heavy quarkonium nonperturbatively.
The polarization of the produced heavy quarkonium from this LP production chain
is determined by the polarization of the  fragmenting parton at the input scale, 
and the dynamics behind the emergence of the heavy quark pair from 
the fragmenting parton, as well as the emergence of the heavy quarkonium 
from the heavy quark pair.

In this whole LP production chain of a heavy quarkonium at high $p_T$, 
no heavy quark pair was produced in the fragmentation process all the way down to 
the input scale $\mu_0\gtrsim 2m_Q$, at which the logarithmic contribution 
in terms of $\ln(\mu_0/(2m_Q))$ is comparable with the corrections 
in powers of $2m_Q/\mu_0$.  More precisely, the heavy quark pair, necessary for the 
production of the heavy quarkonium, is only produced as a part of the input FFs.  
When the available phase space for radiation is much smaller than the mass of heavy quarks,
the polarization of the $J/\psi$ should be very much the same 
as the polarization of the heavy quark pair, which is more or less the same 
as the polarization of the ``final" parton, most often a gluon that splits into the pair. 
Since a virtual gluon of invariant mass much less than its momentum is likely to be 
transversely polarized, the LP QCD factorization formalism for the production naturally 
predicts that $J/\psi$ produced at a high $p_T$ is  transversely polarized,
which is also consistent with the prediction of the NRQCD factorization formalism 
\cite{Brambilla:2010cs}.  However, almost all data on the polarization of high $p_T$ 
$J/\psi$ as well as $\Upsilon$, produced at Tevatron and the LHC energies, do not 
favor the dominance of transverse polarization at the current values of $p_T$, instead, 
they are consistent with no strong polarization.  This is in fact an outstanding puzzle 
whose solution must be crucial for understanding the true mechanism of 
heavy quarkonium production in high energy collisions.

If the NLP term in the QCD factorization formalism in Eq.~(\ref{eq:pqcd_fac0}) is important, 
which seems to be the case \cite{Chao:2012iv,Gong:2013qka,Bodwin:2014gia,Ma:2014svb}, 
heavy quark pairs produced at all distance scales from ${\cal O}(1/p_T)$ to ${\cal O}(1/2m_Q)$
could contribute to the production of heavy quarkonia significantly, and the knowledge of 
heavy quarkonium FFs from a heavy quark pair at the input scale $\mu_0$ is then crucial 
for understanding the polarization of produced heavy quarkonia.  
In the following, we use an example to describe the calculations of these input FFs 
to a polarized $J/\psi$.  More details for fragmentation via other NRQCD states
can be found in Ref.~\cite{Zhang:thesis}.

Like the perturbative NRQCD calculation for the unpolarized heavy quarkonium FFs, 
as discussed in the previous section, we use the same Feynman diagrams 
in Fig.~\ref{fig:fragfunc} for calculating the FFs to a polarized heavy quarkonium, but 
with different spin projection operators to identify the polarization states of 
the produced heavy quark pair (the upper lines in Fig.~\ref{fig:fragfunc}).  
For a heavy quark pair moving in the $+z$-direction, we can write
$p^\mu=p\cdot\hat{n}\, \bar{n}^\mu + p^2/(2p\cdot\hat{n})\,\hat{n}^\mu$, 
with two auxiliary vectors,
$\bar{n}^\mu = (\bar{n}^+, \bar{n}^-, \bar{\bf n}_\perp) = (1,0,{\bf 0}_\perp)$ and
$\hat{n}^\mu = (0,1,{\bf 0}_\perp)$, and we define the polarization vector for a longitudinally 
polarized spin-1 heavy quark pair,
\begin{equation}
\epsilon_{\lambda=0}^\mu 
=\frac{1}{\sqrt{p^2}}\left[
p\cdot\hat{n}\, \bar{n}^\mu - \frac{p^2}{2p\cdot \hat{n}} \hat{n}^\mu 
\right] \, ,
\label{eq:pol_vec_l}
\end{equation}
with $p\cdot \epsilon_0 = 0$ and $\epsilon_0^2=-1$.  From $\epsilon_{\lambda=0}^\mu$, we 
can derive the following spin polarization tensors in this frame, which is 
effectively the S-helicity frame \cite{Lam:1978pu},
\begin{eqnarray}
{\cal P}^{\mu\nu}_L(p) 
&\equiv & \epsilon^{*\mu}_{0}\, \epsilon^{\nu}_{0}
= \frac{1}{p^2}
\left[ p\cdot\hat{n}\, \bar{n}^\mu - \frac{p^2}{2p\cdot \hat{n}} \hat{n}^\mu \right]
\left[ p\cdot\hat{n}\, \bar{n}^\nu - \frac{p^2}{2p\cdot \hat{n}} \hat{n}^\nu \right]
\, ,
\nonumber \\
{\cal P}^{\mu\nu}_T(p) 
&\equiv & \frac{1}{2}
\sum_{\lambda=\pm 1} \epsilon^{*\mu}_{\lambda}\epsilon^{\nu}_{\lambda}
=  \frac{1}{2}
\left[-g^{\mu\nu}+\bar{n}^\mu \hat{n}^\nu + \hat{n}^\mu \bar{n}^\nu \right]
= \frac{1}{2}
\left[{\cal P}^{\mu\nu}(p) -
{\cal P}^{\mu\nu}_L(p) \, ,
\right]\, ,
\label{eq:pol-tensors}
\end{eqnarray}
for producing a longitudinally and transversely polarized spin-1 heavy quark pair
of momentum $p$, respectively. 
The tensor ${\cal P}^{\mu\nu}(p)$ in Eq.~(\ref{eq:pol-tensors}) is defined as
\begin{eqnarray}
{\cal P}^{\mu\nu}(p) 
&=& 
\sum_{\lambda=0,\pm 1} \epsilon^{*\mu}_{\lambda}\epsilon^{\nu}_{\lambda}
= -g^{\mu\nu} + \frac{p^\mu p^\nu}{p^2}\, ,
\label{eq:unpol-tensor}
\end{eqnarray}
which is the polarization tensor for an unpolarized spin-1 heavy quark pair 
of total momentum $p$, which was used for calculating the unpolarized 
input FFs in the last section.

Similar to those unpolarized input FFs, presented in Eqs.~(\ref{eq:frag-v8}), 
(\ref{eq:frag-a8}), (\ref{eq:fragDR-v8}) and (\ref{eq:fragDR-a8}), 
we derive the input FFs to a polarized heavy quarkonium $H$ via a color singlet 
spin-1 NRQCD heavy quark pair, 
by using the polarization tensors in Eq.~(\ref{eq:pol-tensors}), 
\begin{eqnarray}
&& 
{\cal D}^{L,\text{CR}}_{[Q\bar{Q}(v8)]\to [Q\bar{Q}(\CScSa)]\to H}(z, u, v, \mu^2; m_Q)
\nonumber\\
&& {\hskip 0.3 in}
=
\frac{1}{2N_c^2}\frac{\langle {\cal O}^{H}_{[Q\bar{Q}(\CScSa)]}\rangle}{3m_Q}
\Delta_{-}(u, v)
\times
\frac{\alpha_s}{2\pi} \frac{z}{1-z}
\left[
\ln\left(r(z) + 1\right)
- \left(1- \frac{1}{1+r(z)}\right)
\right],
\nonumber\\
&&
{\cal D}^{T,\text{CR}}_{[Q\bar{Q}(v8)]\to [Q\bar{Q}(\CScSa)]\to H}(z, u, v, \mu^2; m_Q)
\nonumber\\
&& {\hskip 0.3 in}
=
\frac{1}{2N_c^2}\frac{\langle {\cal O}^{H}_{[Q\bar{Q}(\CScSa)]}\rangle}{3m_Q}
\Delta_{-}(u, v)
\times
\frac{\alpha_s}{2\pi}z(1-z)\left[
1- \frac{1}{1+r(z)}
\right],
\label{eq:fragpol-v8}
\end{eqnarray}
for the fragmentation of a $[Q\bar{Q}(v8)]$ state, and \cite{Kang:2011mg}
\begin{eqnarray}
&&
{\cal D}^{L,\text{CR}}_{[Q\bar{Q}(a8)]\to [Q\bar{Q}(\CScSa)]\to H}(z, u, v, \mu^2; m_Q)
\nonumber\\
&& {\hskip 0.3 in}
=
\frac{1}{2N_c^2}\frac{\langle {\cal O}^{H}_{[Q\bar{Q}(\CScSa)]}\rangle}{3m_Q}
\Delta_{+}(u, v)
\times
\frac{\alpha_s}{2\pi}z(1-z)
\left[
\ln\left(r(z) + 1\right)
- \left(1- \frac{1}{1+r(z)}\right)
\right],
\nonumber\\
&&
{\cal D}^{T,\text{CR}}_{[Q\bar{Q}(a8)]\to [Q\bar{Q}(\CScSa)]\to H}(z, u, v, \mu^2; m_Q)
\nonumber\\
&& {\hskip 0.3 in}
=
\frac{1}{2N_c^2}\frac{\langle {\cal O}^{H}_{[Q\bar{Q}(\CScSa)]}\rangle}{3m_Q}
\Delta_{+}(u, v)
\times
\frac{\alpha_s}{2\pi}z(1-z)\left[
1- \frac{1}{1+r(z)}
\right],
\label{eq:fragpol-a8}
\end{eqnarray}
for the fragmentation of a $[Q\bar{Q}(a8)]$ state.  
Comparing with those unpolarized input FFs derived in the last section, 
it is clear that the relation, ${\cal D}(p)= 2\,{\cal D}^T(p)+{\cal D}^L(p)$, is satisfied.
Similarly, in a dimensional regularization and $\overline{\rm MS}$ renormalization scheme, 
we have 
\begin{eqnarray}
&&
{\cal D}^{L,\text{DR}}_{[Q\bar{Q}(v8)]\to [Q\bar{Q}(\CScSa)]\to H}(z, u, v, \mu^2; m_Q)
\nonumber\\
&& {\hskip 0.3in}
=
\frac{1}{2N_c^2}\frac{\langle {\cal O}^{H}_{[Q\bar{Q}(\CScSa)]}\rangle}{3m_Q}
\Delta_{-}(u, v)
\times
\frac{\alpha_s}{2\pi} \frac{z}{1-z}
\left[
\ln\left(\frac{\mu^2}{4(1-z)^2m_Q^2}\right)
- 1 \right],
\nonumber\\
&&
{\cal D}^{T,\text{DR}}_{[Q\bar{Q}(v8)]\to [Q\bar{Q}(\CScSa)]\to H}(z, u, v, \mu^2; m_Q)
\nonumber\\
&& {\hskip 0.3in}
=
\frac{1}{2N_c^2}\frac{\langle {\cal O}^{H}_{[Q\bar{Q}(\CScSa)]}\rangle}{3m_Q}
\Delta_{-}(u, v)
\times
\frac{\alpha_s}{2\pi}z(1-z),
\label{eq:fragpolDR-v8}
\end{eqnarray}
for the fragmentation of a $[Q\bar{Q}(v8)]$ state, and 
\begin{eqnarray}
&&
{\cal D}^{L,\text{DR}}_{[Q\bar{Q}(a8)]\to [Q\bar{Q}(\CScSa)]\to H}(z, u, v, \mu^2; m_Q)
\nonumber\\
&& {\hskip 0.3 in}
=
\frac{1}{2N_c^2}\frac{\langle {\cal O}^{H}_{[Q\bar{Q}(\CScSa)]}\rangle}{3m_Q}
\Delta_{+}(u, v)
\times
\frac{\alpha_s}{2\pi}z(1-z)
\left[
\ln\left(\frac{\mu^2}{4(1-z)^2m_Q^2}\right)
- 3 \right],
\nonumber\\
&&
{\cal D}^{T,\text{DR}}_{[Q\bar{Q}(a8)]\to [Q\bar{Q}(\CScSa)]\to H}(z, u, v, \mu^2; m_Q)
\nonumber\\
&& {\hskip 0.3 in}
=
\frac{1}{2N_c^2}\frac{\langle {\cal O}^{H}_{[Q\bar{Q}(\CScSa)]}\rangle}{3m_Q}
\Delta_{+}(u, v)
\times
\frac{\alpha_s}{2\pi}z(1-z),
\label{eq:fragpolDR-a8}
\end{eqnarray}
for the fragmentation of a $[Q\bar{Q}(a8)]$ state.  
In Eqs.~(\ref{eq:fragpol-v8}), (\ref{eq:fragpol-a8}), (\ref{eq:fragpolDR-v8}), and 
(\ref{eq:fragpolDR-a8}) above,
we have assumed that the NRQCD LDME for an unpolarized spin-1 heavy quark pair 
to an unpolarized heavy quarkonium $H$, $\langle {\cal O}^{H}_{[Q\bar{Q}(\CScSa)]}\rangle$,
is the same as that for a polarized spin-1 heavy quark pair 
to a polarized heavy quarkonium $H$ in the same polarization state.

In the polarized input FFs in Eqs.~(\ref{eq:fragpol-v8}), (\ref{eq:fragpol-a8}), 
(\ref{eq:fragpolDR-v8}), and (\ref{eq:fragpolDR-a8}), it is interesting to note 
the following feature: 
the FFs to a longitudinally polarized heavy quark pair are enhanced by
a logarithmic term, $\ln(r(z)+1) \approx \ln(1/(1-z)^2) + ...$, as $z\to 1$, 
while those to a transversely polarized pair are not.
This is a natural result of the UV power counting because only 
the $p\cdot\hat{n}\, \bar{n}^\mu$ term of the longitudinal polarization vector, 
$\epsilon_0^{\mu}$ in Eq.~(\ref{eq:pol_vec_l}), 
which leads to an equivalent spin contraction $\gamma\cdot p$ 
for the produced heavy quark pair, picks up the leading 
logarithmic UV divergence of the diagrams in Fig.~\ref{fig:fragfunc}
when the fragmenting heavy quark pair is in a vector or axial vector spin state
(contracted by $\gamma\cdot \hat{n}$ or $\gamma\cdot \hat{n}\gamma_5$).
While the UV divergence, when the transverse momentum of the radiated gluon 
$k_\perp^2 \to \infty$, is renormalized leading to the logarithmic factorization scale 
$\mu^2$-dependence of the FFs, the opposite limit when $k_\perp^2\to 0$ 
corresponding to $z\to 1$ gives the logarithmic enhancement of the FFs
to a longitudinally polarized heavy quark pair.
That is, a perturbatively produced color octet heavy quark pair, 
in either a vector or an axial vector spin state, 
is more likely to fragment into a longitudinally polarized color singlet spin-1 NRQCD 
heavy quark pair when the momentum fraction $z$, carried by the NRQCD heavy quark pair,
is large. 

From the QCD factorization formalism in Eq.~(\ref{eq:pqcd_fac0}), the hadronic 
cross section for heavy quarkonium production at large $p_T$ is proportional to 
a convolution of two PDFs and one FF for each partonic scattering channel.  
With two steeply falling PDFs as a function of parton momentum fraction $x$, the 
hadronic production cross section is dominated by the phase space where the 
momentum fractions $x$ of colliding partons are small while the produced outgoing 
hadron momentum fraction $z$ is large \cite{Berger:2001wr}.  That is, the heavy quarkonium
produced by the fragmentation of a perturbatively produced heavy quark pair via a 
color singlet and spin-1 NRQCD heavy quark pair is likely to be longitudinally polarized.
To demonstrate this feature quantitatively, we define the polarization parameter,
\begin{equation}
\alpha(p) \equiv 
\frac{\sigma^T_{A+B\to J/\psi(p)} - \sigma^L_{A+B\to J/\psi(p)}}
       {\sigma^T_{A+B\to J/\psi(p)} + \sigma^L_{A+B\to J/\psi(p)}}\ , 
\label{eq:pol-para}
\end{equation}
where $\sigma^T_{A+B\to J/\psi(p)}$ and $\sigma^L_{A+B\to J/\psi(p)}$ are the cross sections 
for producing a transversely and a longitudinally polarized $J/\psi$ of momentum $p$, respectively, 
and are given by the same factorized expressions in Eq.~(\ref{eq:singlet-lo}) with
the FFs replaced by corresponding FFs to a transversely (or longitudinally) polarized $J/\psi$
in Eq.~(\ref{eq:fragpol-v8}) (or in Eq.~(\ref{eq:fragpol-a8})).
In the bottom panel of Fig.~\ref{fig:cutoff}, we have plotted the polarization parameter $\alpha(p)$ 
for $J/\psi$ production as a function of its $p_T$.    
It is clear from Fig.~\ref{fig:cutoff} that most $J/\psi$'s produced
through spin-1 and color singlet heavy quark pairs are longitudinally polarized  \cite{Kang:2011mg}.
Like the cross section shown in Fig.~\ref{fig:cutoff}, the polarization parameter $\alpha$, 
calculated by using the QCD factorization formalism at the LO and with the calculated 
FFs, completely reproduces the NLO CSM calculation in Ref.~\cite{Gong:2008sn}.  
Having the complete set of input FFs to a polarized heavy quarkonium \cite{Zhang:thesis}, 
which are universal, it should be straightforward in terms of QCD factorization formalism in 
Eq.~(\ref{eq:pqcd_fac0}) to evaluate the production rate for both transversely and longitudinally 
polarized heavy quarkonia in high energy hadron-hadron, hadron-lepton, and lepton-lepton 
scatterings, and to test the QCD factorization formalism and our understanding of 
the mechanism responsible for the heavy quarkonium production, 
which we leave for future work.

There have been many proposals to resolve the polarization puzzle of 
heavy quarkonum polarization \cite{Brambilla:2010cs}, and many of them are  
within the NRQCD factorization approach \cite{Chao:2012iv,Gong:2013qka,Bodwin:2014gia}.
By adjusting the value of NRQCD LDMEs so that the two leading power production channels, 
via $[Q\bar{Q}(\COcSa)]$ and $[Q\bar{Q}(\COcPj)]$ states, 
which likely produce the transversely polarized $J/\psi$, are canceled between them, 
it is possible to leave the production dominated by the channel with an unpolarized 
$[Q\bar{Q}(\COaSz)]$ state.  As demonstrated in Ref.~\cite{Bodwin:2014gia}, 
the contribution from this channel for the relevant $p_T$ region cannot be reproduced 
by the QCD factorized fragmentation restricted to  LP.  
Instead, as demonstrated in Ref.~\cite{Ma:2014svb}, 
the contribution from the $[Q\bar{Q}(\COaSz)]$ channel from
the state of the art NLO NRQCD calculation is completely reproduced 
by the LO contribution from QCD factorization in Eq.~(\ref{eq:pqcd_fac0}), 
evaluated with heavy quarkonium FFs calculated in NRQCD, 
and is found to be dominated by the NLP contribution for the most relevant 
$p_T$ range of the existing data.  It seems likely, then, that the NLP contribution
to heavy quarkonium production is crucial for resolving the outstanding puzzles of 
heavy quarkonium polarization at the current collision energies.

In terms of the QCD factorization formalism in Eq.~(\ref{eq:pqcd_fac0}), there are two 
major sources of NLP contributions that could generate heavy quarkonium polarization 
different from that at the LP.  One is directly from the NLP term of the factorization 
formalism, or more specifically, from the heavy quark pair FFs to a heavy quarkonium, 
and the other is indirectly from the LP term due to the NLP corrections to the evolution
equations of the single parton FFs to a heavy quarkonium \cite{Kang:2014tta}.  

The direct contribution to heavy quarkonium polarization should
come from the knowledge of input FFs to a polarized heavy quarkonium
since the partonic hard parts are insensitive to the details of the hadronic states produced. 
Like the color singlet contribution discussed in this section, longitudinally 
polarized heavy quarkonia at large $p_T$ can be naturally estimated 
using the model FFs of \cite{Ma:2013yla,Ma:2014eja,Zhang:thesis}, in this direct NLP contribution.  

As required by the consistency of QCD factorization at NLP accuracy, 
which was pointed out in Ref.~\cite{Kang:2014tta}, the DGLAP evolution equation
for the factorization scale dependence of the single-parton FFs to a heavy quarkonium 
needs to be modified to include an NLP correction.  This modification effectively takes into 
account the power suppressed contribution to the evolution of single parton 
FFs, in which a single parton evolves into a heavy quark pair, in addition to its evolution to
other single partons at  LP.   The pair subsequently evolves into a heavy quarkonium
via the heavy quark pair FFs. While the LP evolution of the single parton FFs leads to 
a dominance of transverse polarization, the NLP correction to the evolution of single parton 
FFs could lead to more longitudinally polarized heavy quarkonia.
It is clear that both the direct and the indirect NLP contributions to 
heavy quarkonium production from the QCD factorization formalism 
in Eq.~(\ref{eq:pqcd_fac0}) reduce the dominance of transversely polarized heavy quarkonia, 
as predicted by the purely LP fragmentation contribution to heavy quarkonium production.
It is therefore critically important to evaluate the production rate of polarized heavy quarkonia
in high energy scattering in terms of the QCD factorization formalism in Eq.~(\ref{eq:pqcd_fac0}) 
using the heavy quarkonium FFs calculated in the NRQCD factorization approach 
\cite{Ma:2013yla,Ma:2014eja,Zhang:thesis}.

\newpage
\section{Summary and conclusions}
\label{sec:summary}

We have calculated, in terms of the QCD collinear factorization formalism in Eq.~(\ref{eq:pqcd_fac0}), 
a complete set of short-distance partonic hard parts at ${\cal O}(\alpha_s^3)$ 
for the NLP contribution to hadronic heavy quarkonium production 
at large transverse momentum $p_T$ at collider energies.  
This new factorization formalism organizes the 
production cross section of heavy quarkonia at large $p_T$ in terms of 
a power expansion of $1/p_T$, with both the LP and NLP contributions
factorized into convolutions of perturbatively calculable 
short-distance partonic hard parts and the universal, but nonperturbative 
heavy quarkonium FFs \cite{Kang:2014tta}.  Like all QCD factorization formalisms, 
the short-distance partonic hard parts are insensitive to the details of the hadrons
produced.  The short-distance hard parts at LP are effectively the same as the 
hard parts for LP hadronic production of light hadrons, and are available in the literature
for both LO and NLO at ${\cal O}(\alpha_s^2)$ and ${\cal O}(\alpha_s^3)$, respectively 
\cite{Aversa:1988vb}.  Our calculated LO partonic hard parts to the NLP contribution at 
${\cal O}(\alpha_s^3)$, in principle, depend on the separation of LP 
from the NLP contribution to the cross section from the same partonic scattering diagrams.
In practice, as discussed in Sec.~\ref{sec:hardparts}, we introduce a gluon contact
term to remove the LP contribution analytically in our calculation at ${\cal O}(\alpha_s^3)$.

With the short-distance partonic hard parts calculated in this paper and 
the evolution kernels for the scale dependence of FFs calculated in Ref.~\cite{Kang:2014tta}, 
the predictive power of the QCD factorization formalism for heavy quarkonium production
still depends on our knowledge of heavy quarkonium FFs at an input scale $\mu_0$.  
Because of the large heavy quark mass, $m_Q\gg \Lambda_{\rm QCD}$, and the clear
separation of momentum scales between the perturbative scales, $\mu_0\gtrsim m_Q$, 
and the nonperturbative scales of the input FFs, such as heavy quark momentum $\sim m_Q v$, 
and energy $\sim m_Q v^2$, we have proposed, as a conjecture or  model, to use NRQCD 
factorization  to evaluate the heavy quarkonium FFs at the input scale.  Then, 
the large number of unknown heavy quarkonium FFs from either a single parton 
or a heavy quark pair can be factorized into perturbatively calculable functional dependence
on momentum fractions, $z$, $\zeta_1$ and $\zeta_2$, at the NRQCD factorization 
scale $\mu_\Lambda\sim m_Q \sim {\cal O}(\mu_0)$, which can be combined with a few universal 
NRQCD LDMEs organized in terms of their effective powers in heavy quark velocity $v$. 
This increases the predictive power of the QCD factorization formalism, as well
as its testability.  Since the QCD factorization of the production cross section and the NRQCD
factorization of the universal FFs at the input scale are two independent factorizations, 
using different expansion parameters and power counting, we can improve overall
predictive power or accuracy on the production cross section by increasing the perturbative 
accuracy of each perturbative expansion. 
 
If the NRQCD factorization for the universal input FFs is proved to be valid, 
the QCD factorization approach with calculated FFs in NRQCD is effectively equal 
to the NRQCD factorization for the first two powers of $1/p_T$ expansion of 
the production cross section, although the QCD factorization and NRQCD factorization 
organize their perturbative expansions of the cross section differently.  
The QCD factorization includes all order resummation of $\ln(p_T^2/m_Q^2)$-type 
large logarithmic contribution at high $p_T$, and is more suited for heavy quarkonium 
production in the high $p_T$ region, while the NRQCD factorization, including the explicit 
heavy quark mass dependence, is better for the production at $p_T \gtrsim m_Q$.  
We proposed a matching equation in Eq.~(\ref{eq:matching}) to expand the coverage 
of the factorization formalism for heavy quarkonium production at collider energies.

Understanding the polarization of produced heavy quarkonia in high energy scattering
is a major challenge for the NRQCD factorization formalism.  We demonstrated that 
the QCD collinear factorization including the NLP contribution associated with short-distance heavy pair production has potentially both 
direct and indirect ways to suppress the dominance of transverse polarization predicted
by the LP fragmentation contribution.  Since short-distance coefficients are insensitive
to  hadron properties, the universal FFs at the input scale are largely responsible 
for the polarization of produced heavy quarkonia, while the NLP corrections 
to the evolution equations of single parton FFs may also be very important to reduce the 
dominance of transverse polarization.  The input FFs to a polarized heavy quarkonium
state, both longitudinal and transverse, are now calculated in the NRQCD factorization 
approach \cite{Ma:2013yla,Ma:2014eja,Zhang:thesis}.  With the partonic short-distance hard parts calculated in
this paper, the evolution kernels of heavy quarkonium FFs calculated in 
Ref.~\cite{Kang:2014tta}, and input FFs to a polarized heavy quarkonium, we are now
in a position to evaluate, consistently, QCD predictions for the polarization of heavy 
quarkonia produced in high energy scatterings.  By calculating partonic hard parts 
for heavy quarkonium production in $e^+e^-$ and lepton-hadron collisions, and using 
the existing universal evolution kernels and input FFs, it is completely possible to 
perform a QCD global analysis of all data on heavy quarkonium production to test our 
understanding on how heavy quarkonia are really produced, forty years since the 
first heavy quarkonium, $J/\psi$, was discovered \cite{Aubert:1974js,Augustin:1974xw}.

\section*{Acknowledgments}

We thank G.T. Bodwin, E. Braaten and H. Zhang for helpful discussions.  
This work was supported in part by the U. S. Department of Energy under Contracts No.~DE-AC02-06NA25396 and No. DE-AC02-98CH10886, and Grant No.~DE-FG02-93ER-40762, 
and the National Science Foundation under Grants No.~PHY-0969739 and No. PHY-1316617.

\newpage
\appendix
\section{Partonic hard parts}
\label{app:hardparts}

In this appendix, we summarize the partonic hard parts for the NLP contribution to heavy quarkonium production from all partonic scattering channels, 
$i(p_1)+j(p_2)\to [Q\bar{Q}(\kappa)](p)+k(p_3)$, 
for hadronic collisions.  We define the hard part, $H_{ij\to [Q\bar{Q}(\kappa)]}$, 
in terms of the invariant partonic cross section,
\begin{equation}
E_p \frac{d\hat{\sigma}_{ij\to [Q\bar{Q}(\kappa)]}}{d^3p}
= \frac{4\pi\alpha_s^3}{\hat{s} }\, \frac{1}{\bar{u} u \bar{v} v}\,
H_{ij\to [Q\bar{Q}(\kappa)]} \, 
\delta\left(\hat{s}+\hat{t}+\hat{u}\right)\, ,
\end{equation}
with Mandelstam variables defined as
\begin{align}
\hat{s}=(p_1+p_2)^2=(p+p_3)^2,\nonumber\\
\hat{t}=(p_2-p_3)^2=(p-p_1)^2,\nonumber\\
\hat{u}=(p_1-p_3)^2=(p-p_2)^2.\nonumber
\end{align}
As discussed in Sec.~\ref{sec:hardparts}, the partonic hard parts at NLP depend on the subtraction of the LP contribution from the partonic scattering, and the corresponding choice of the regularization.  The following results are calculated by using the gluon contact term to remove the LP contribution at this order analytically.  The dependence on the auxiliary vector $\hat{n}^\mu$, which defines the gluon contact term, is kept explicit, in the form of $p\cdot \hat{n}\equiv p^+$ for any momentum vector $p$.  More discussion on the definition and the choice of $\hat{n}^\mu$ can be found in Sec.~\ref{sec:hardparts}.

\noindent{\it 1) quark-antiquark scattering:}
\begin{align}
H_{q\bar{q}\to\left[Q\bar{Q}(a1)\right]g}=&\frac{N_c^2-1}{2 N_c^3}\frac{\hat{t}^2+\hat{u}^2}{\hat{s}^3},\\
H_{q\bar{q}\to\left[Q\bar{Q}(v1)\right]g}=&\frac{N_c^2-1}{2 N_c^3}\frac{\hat{t}^2+\hat{u}^2}{\hat{s}^3}\zeta_1 \zeta_2,\\
H_{q\bar{q}\to\left[Q\bar{Q}(a8)\right]g}=&\frac{N_c^2-1}{4 N_c}\frac{\hat{t}^2+\hat{u}^2}{\hat{s}^3}\left(\frac{N_c^2-4}{N_c^2}+\zeta_1 \zeta_2\right),\\
H_{q\bar{q}\to\left[Q\bar{Q}(v8)\right]g}=&\frac{N_c^2-1}{4 N_c}\frac{\hat{t}^2+\hat{u}^2}{\hat{s}^3}\left(\frac{N_c^2-4}{N_c^2}+\zeta_1
\zeta_2\right)\zeta_1 \zeta_2 \nonumber\\
&-\frac{N_c^2-1}{4 N_c}\left\{2\left(\frac{1}{\hat{u}}\frac{p_1^+}{p^+}+\frac{1}{\hat{t}}\frac{p_2^+}{p^+}\right)
\left(\frac{\hat{t}^2+\hat{u}^2}{\hat{s}^2}-\frac{1}{N_c^2}\right)\left(1-\zeta_1^2\right)\left(1-\zeta_2^2\right)\right. \nonumber \\
&+\left. \frac{1}{2}\left[\left(\frac{1}{\hat{t}}\frac{p_1^+}{p^+}-\frac{1}{\hat{u}}\frac{p_2^+}{p^+}\right)\frac{\hat{t}^2+\hat{u}^2}{\hat{s}^2}-
\left(\frac{1}{\hat{t}}-\frac{1}{\hat{u}}\right)\right] \right.\nonumber\\
&\times \left.\left[\frac{\hat{t}-\hat{u}}{\hat{s}}\left(\zeta_1^2+\zeta_2^2-2\zeta_1^2\zeta_2^2\right) + \frac{N_c^2-4}{N_c^2}
\left(\zeta_1+\zeta_2\right) \left(1-\zeta_1\zeta_2\right)
\right]\right\}\,.\label{eq:qqtoQQv8g}
\end{align}
In  Eq.~\eqref{eq:qqtoQQv8g}, the term proportional to $\zeta_1 \zeta_2$ comes from the contribution of (a)+(b) in Fig.~\ref{fig:qq2QQg},
and the term proportional to $\left(\zeta_1+\zeta_2\right) \left(1-\zeta_1\zeta_2\right)$ comes from the contribution of interference between (a)+(b) and
(c)+(d)+(e) in Fig.~\ref{fig:qq2QQg}. It is straightforward to check charge conjugation symmetry
\begin{align}
H_{q\bar{q}\to\left[Q\bar{Q}(\kappa)\right]g}(p_1, p_2, p_3, \zeta_1, \zeta_2)=H_{q\bar{q}\to\left[Q\bar{Q}(\kappa)\right]g}(p_2, p_1, p_3, -\zeta_1, -\zeta_2).
\end{align}
Results for $\bar{q}q$ initial states can be obtained from results of $q\bar{q}$ initial states by exchanging $p_1$ and $p_2$,
\begin{align}
H_{\bar{q}q\to\left[Q\bar{Q}(\kappa)\right]g}(p_1, p_2, p_3, \zeta_1, \zeta_2)&=H_{q\bar{q}\to\left[Q\bar{Q}(\kappa)\right]g}(p_2, p_1, p_3, \zeta_1, \zeta_2)\nonumber\\
&=H_{q\bar{q}\to\left[Q\bar{Q}(\kappa)\right]g}(p_1, p_2, p_3, -\zeta_1, -\zeta_2).
\end{align}

\noindent{\it 2) quark-gluon and gluon-quark scattering:}
Similarly, results for $q g$, $\bar{q} g$, $g q$ and $g \bar{q}$ 
initial states are given by the following crossing relationships, 
\begin{align}
H_{q g\to\left[Q\bar{Q}(\kappa)\right]q}(p_1, p_2, p_3, \zeta_1, \zeta_2)&=-\frac{N_c}{N_c^2-1}H_{q\bar{q}\to\left[Q\bar{Q}(\kappa)\right]g}(p_1, -p_3, -p_2, \zeta_1, \zeta_2),\\
H_{\bar{q} g\to\left[Q\bar{Q}(\kappa)\right]\bar{q}}(p_1, p_2, p_3, \zeta_1, \zeta_2)&=-\frac{N_c}{N_c^2-1}H_{q\bar{q}\to\left[Q\bar{Q}(\kappa)\right]g}(p_1, -p_3, -p_2, -\zeta_1, -\zeta_2),\\
H_{g q\to\left[Q\bar{Q}(\kappa)\right]q}(p_1, p_2, p_3, \zeta_1, \zeta_2)&=-\frac{N_c}{N_c^2-1}H_{q\bar{q}\to\left[Q\bar{Q}(\kappa)\right]g}(p_2, -p_3, -p_1, \zeta_1, \zeta_2),\\
H_{g \bar{q}\to\left[Q\bar{Q}(\kappa)\right]\bar{q}}(p_1, p_2, p_3, \zeta_1, \zeta_2)&=-\frac{N_c}{N_c^2-1}H_{q\bar{q}\to\left[Q\bar{Q}(\kappa)\right]g}(p_2, -p_3, -p_1, -\zeta_1, -\zeta_2),
\end{align}
where the color factor accounts for differences of averaging over initial color states. We give the results for $g q$ initial states explicitly,
\begin{align}
H_{gq\to\left[Q\bar{Q}(a1)\right]q}=&-\frac{1}{2 N_c^2}\frac{\hat{u}^2+\hat{s}^2}{\hat{t}^3},\\
H_{gq\to\left[Q\bar{Q}(v1)\right]q}=&-\frac{1}{2 N_c^2}\frac{\hat{u}^2+\hat{s}^2}{\hat{t}^3}\zeta_1 \zeta_2,\\
H_{gq\to\left[Q\bar{Q}(a8)\right]q}=&-\frac{1}{4}\frac{\hat{u}^2+\hat{s}^2}{\hat{t}^3}\left(\frac{N_c^2-4}{N_c^2}+\zeta_1 \zeta_2\right),\\
H_{gq\to\left[Q\bar{Q}(v8)\right]q}=&-\frac{1}{4}\frac{\hat{u}^2+\hat{s}^2}{\hat{t}^3}\left(\frac{N_c^2-4}{N_c^2}+\zeta_1
\zeta_2\right)\zeta_1 \zeta_2 \nonumber\\
&+\frac{1}{4}\left\{2\left(\frac{1}{\hat{s}}\frac{p_2^+}{p^+}-\frac{1}{\hat{u}}\frac{p_3^+}{p^+}\right)
\left(\frac{\hat{u}^2+\hat{s}^2}{\hat{t}^2}-\frac{1}{N_c^2}\right)\left(1-\zeta_1^2\right)\left(1-\zeta_2^2\right)\right. \nonumber \\
&+\left. \frac{1}{2}\left[\left(\frac{1}{\hat{u}}\frac{p_2^+}{p^+}+\frac{1}{\hat{s}}\frac{p_3^+}{p^+}\right)\frac{\hat{u}^2+\hat{s}^2}{\hat{t}^2}-
\left(\frac{1}{\hat{u}}-\frac{1}{\hat{s}}\right)\right] \right.\nonumber\\
&\times \left.\left[\frac{\hat{u}-\hat{s}}{\hat{t}}\left(\zeta_1^2+\zeta_2^2-2\zeta_1^2\zeta_2^2\right) + \frac{N_c^2-4}{N_c^2}
\left(\zeta_1+\zeta_2\right) \left(1-\zeta_1\zeta_2\right) \right]\right\}\,.
\end{align}

\noindent{\it 3) gluon-gluon scattering:}
\begin{align}
H_{gg\to\left[Q\bar{Q}(a1)\right]g}=&\frac{2}{N_c^2-1}\frac{S_2^4}{S_3^3},\\
H_{gg\to\left[Q\bar{Q}(v1)\right]g}=&\frac{2}{N_c^2-1}\frac{S_2^4}{S_3^3}\zeta_1 \zeta_2,\\
H_{gg\to\left[Q\bar{Q}(a8)\right]g}=&\frac{N_c^2}{N_c^2-1}\frac{S_2^4}{S_3^3}\left[\frac{N_c^2-4}{N_c^2}+
\left(1-5\frac{S_3^2}{S_2^3}\right)\zeta_1 \zeta_2\right],\\
H_{gg\to\left[Q\bar{Q}(v8)\right]g}=&\frac{N_c^2}{N_c^2-1}\frac{S_2^4}{S_3^3}\left[\frac{N_c^2-4}{N_c^2}+
\left(1-5\frac{S_3^2}{S_2^3}\right)\zeta_1 \zeta_2\right]\zeta_1 \zeta_2 \nonumber\\
&-\frac{N_c^2}{N_c^2-1}\frac{S_2}{S_3^2}\left\{2\left(\hat{t}^3\frac{p_1^+}{p^+}+\hat{u}^3\frac{p_2^+}{p^+}-\hat{s}^3\frac{p_3^+}{p^+}\right)
\left(1-\zeta_1^2\right)\left(1-\zeta_2^2\right)\right. \nonumber \\
&+\left. \frac{1}{2}\left[-3\left(\hat{t}^3\frac{p_1^+}{p^+}+\hat{u}^3\frac{p_2^+}{p^+}-\hat{s}^3\frac{p_3^+}{p^+}\right)
+\frac{2S_2^2}{S_3}\left(\hat{t}^2\frac{p_1^+}{p^+}+\hat{u}^2\frac{p_2^+}{p^+}-\hat{s}^2\frac{p_3^+}{p^+}\right) -4S_3\right] \right.\nonumber\\
&\times \left(\zeta_1^2+\zeta_2^2-2\zeta_1^2\zeta_2^2\right)
\bigg\}\,,
\end{align}
where we have used the method described in Ref.~\cite{Ma:2012ex} to write these expressions in symmetric form, with symmetric variables defined as
\begin{align}
S_2&=-\hat{s}\hat{t}-\hat{t}\hat{u}-\hat{u}\hat{s},\nonumber\\
S_3&=\hat{s}\hat{t}\hat{u}.
\end{align}

\bibliographystyle{h-physrev5} 
\bibliography{./references/references}{}

\end{document}